\documentclass[aps,pre,amsmath,amssymb,nofootinbib,superscriptaddress,twocolumn]{revtex4-1}
\usepackage{graphicx}

\usepackage{graphicx}
\usepackage{amsmath,pgfmath,pgffor}
\usepackage[usenames, dvipsnames]{color}
\usepackage{indentfirst}
\usepackage{hyperref}

\usepackage[T1]{fontenc}
\usepackage[english, american]{babel}
\usepackage{amsmath}
\usepackage[utf8]{inputenc}

\usepackage{subcaption}

\usepackage{makeidx}
\usepackage{bigints}

\usepackage{colortbl}

\usepackage{feyn}

\usepackage{color}

\hypersetup{%
    pdfborder = {0 0 0},
    colorlinks,
    citecolor=magenta,
    linkcolor=blue,
}
\DeclareMathOperator\erfc{erfc}

\begin{document}

\title{Well-mixed Lotka-Volterra model with random strongly competitive interactions}
\author{Giulia Garcia Lorenzana} 
\affiliation{Laboratoire de Physique de l'\'Ecole Normale Supérieure, Université PSL, CNRS, Sorbonne Université, Université Paris-Diderot, Paris, France}
\affiliation{Laboratoire Matière et Systèmes Complexes (MSC), Université de Paris, CNRS, 75013 Paris, France}
\author{Ada Altieri}
\affiliation{Laboratoire Matière et Systèmes Complexes (MSC), Université de Paris, CNRS, 75013 Paris, France}

\begin{abstract}
The random Lotka-Volterra model is widely used to describe the dynamical and thermodynamic features of ecological communities.
In this work, we consider random symmetric interactions between species and analyze the strongly competitive interaction case. We investigate different scalings for the distribution of the interactions with the number of species and try to bridge the gap with previous works.
%Different scalings for the distribution of the interactions with the number of species have been considered in the literature and here we will try to bridge this conceptual gap.
Our results show two different behaviors for the mean abundance at zero and finite temperature respectively, with a continuous crossover between the two. We confirm and extend previous results obtained for weak interactions: at zero temperature, even in the strong competitive interaction limit, the system is in a multiple-equilibria phase, whereas at finite temperature only a unique stable equilibrium can exist. 
Finally, we establish the qualitative phase diagrams and compare the species abundance distributions in the two cases.

\end{abstract}

\maketitle

\section{Introduction}

The incredible biodiversity that characterizes natural ecosystems has attracted ecologists for a long time. 
Also, investigating biotic and abiotic interactions along with the determination of their strength is a central topic in ecology given their key role in shaping community stability and spatio-temporal processes
 \cite{Vazquez2012, Bunin2021}.

On a theoretical level, modeling the interactions between many different components -- from bacteria in a microbial community to plant-pollinator impact in a forest to birds in starling flocks -- can get complicated quickly. 
Given the presence of many entities interacting in myriad ways, in the last decades physicists have started drawing parallels with complex and disordered systems. 
As yet, there is no single, well-established theory allowing us to understand such systems and to integrate the plethora of empirical data coming from a growing number of controlled experiments. In particular, a quantitative general framework able to discriminate between the \emph{niche theory} \cite{Tilman2004niche, gupta2021}– based on a limited number of niches, each occupied by a single species according to the competitive exclusion principle \cite{Hardin1960} – and \emph{neutral models} \cite{azaele2016} -- in which species are treated as equivalent -- is still missing. Mechanisms according to which individuals shape and re-organize their \emph{niche} have been studied both at equilibrium and in out-of-equilibrium settings, using for instance time series and ecological successions principles. 
Other key ecological questions generally include: i) relaxation either to a single equilibrium or to a multistability regime; ii) the correct definition of \emph{ecosystem diversity}, i.e. the number of surviving species that characterize each of these equilibria; iii) the typical behavior of fluctuations and functional responses to external perturbations \cite{Arnoldi2018, Roy2020, Altieri2021}; iv) interplay between stochastic and deterministic processes and how community diversity and variability are related to them \cite{Zaoli2021}; v) appearance of chaotic dynamics and periodic cycles to be experimentally measured in real ecosystems \cite{Beninca2008, Pearce2020}.

Addressing these questions is notoriously challenging. However, remarkable advances have been done starting from minimal models and employing Random Matrix Theory and dynamical systems formalism \cite{Tilman1982, van2015, Gibbs2018, Fisher2014, Barbier2018, Sidhom2020, Bunin2016, bunin2017}. 

One of the most employed and most successful frameworks in theoretical ecology is provided by the generalized Lotka Volterra equations, which were proposed one century ago to describe the dynamics inside an ecological community \cite{Lotka1920, volterra1926, MacArthur1955, May2007}. Thanks to their extremely general setting, they have led to applications in many other fields, such as economics, immunology, genetics, and evolutionary game theory \cite{moran2019, Goodwin1990, Behn1992, Bomze1995}.
In the following, we will focus on their mean-field disordered version describing the evolution of a large well-mixed ecosystem with randomly interacting species. Interestingly, in the limit of a large number of components, new and powerful ingredients come into play. 

As it is often the case in statistical physics, despite the lack of information on the underlying microscopical structure, this model turns out to be appropriate to tackle the staggering complexity of ecological systems \cite{Barbier2018, hu2021}. It shows the clear advantage of being exactly solvable and it can reproduce salient features of community dynamics, such as collective behaviors and emergent patterns in terms of phase transitions between different increasingly complex regimes. Notably, critical glassy phases have recently been shown in the Lotka-Volterra model with symmetric random interactions between species \cite{bunin2017, biroli2018, altieri2021a}.

Most of the results in the literature thus far have been obtained in the weak interaction case. Preliminary results in the opposite direction (and with non-symmetric interactions) were shown in \cite{Kessler2015} -- although with a relatively small number of species -- and in \cite{Baron2019, hu2021} pointing out the appearance of a disordered phase with sudden turnovers of the dominant species.
Also, power-law distributions of the species abundances and chaotic dynamics have been experimentally observed in planktonic ecosystems \cite{De2015, SerGiacomi2018, Beninca2008} suggesting that such behaviors can be driven by strong interactions.
Moreover, the hypothesis of some combination of many weak interactions and few strong ones has been proposed to explain the emergence of directional ecological dynamics \cite{Bunin2021}.

In this paper, we aim to complete and extend the analysis by investigating emergent phase transitions for a large number of species and strong symmetric interactions. In this regime, we expect completely different features from the weakly interacting case. 

When the interactions are mediated by the environment, as it is usually the case in microbial systems, the strength of the interactions can be continuously modified by environmental changes, such as the concentration of nutrients \cite{ratzke2020}. 
Studying the strongly interacting limit could therefore be relevant to predict the behavior of an ecosystem in the case of strong environmental modifications (see for instance \cite{hu2021} for a controlled experimental setting). 
The growing interaction strength can also have an evolutionary origin in the case of active competition, in which two competing microbial strains directly harm each other, \emph{i.e.} through the production of toxins or antibiotics \cite{bauer2018}.

The article is organized as follows: in Sec. \ref{sec:model} we will introduce the mathematical model and discuss different possibilities for the scaling of the couplings as a function of the number of species in the pool; in Sec. \ref{sec:literatureresults} we will recap the main results obtained for this model from a thermodynamic perspective using in particular disordered system techniques.
Then in Sec. \ref{sec:zeroT} and Sec. \ref{sec:finiteT} we will enter into the details of our work and analyze the zero-temperature and the finite-temperature regimes. We will present our results in these scenarios in terms of the phase diagrams, the typical scalings for the average abundance as well as the resulting probability distributions. In Sec. \ref{sec:conclusions} we conclude by presenting also some perspectives for future research on this topic.

\section{The model}
\label{sec:model}

We will consider a well-mixed ecological community model: all individuals from all species can interact, and their presence/absence does not depend on the specific location, in good agreement with observations in the oceanic plankton \cite{Baron2019}.

The dynamical equations for the species abundances $N_i$ \cite{biroli2018, altieri2021a} -- where the index $i$ runs over the $S$ different species -- are defined by
\begin{equation}
%	\label{eq:lotkavolterra}
	\frac{d N_i}{dt}= -N_i\left(\nabla_{N_i}V_i(N_i)+\sum_{j\neq i}\alpha_{ij}N_j\right)+\eta_i(t)+m_i \  .
	\label{dynamicaleq}
\end{equation}
 $\alpha_{ij}$ is a random interaction matrix, $\eta_i$ is a white noise and $m_i$ is an immigration rate that couples the community to a regional pool of species, ensuring that all species are present. The immigration rate is typically assumed to be small in order not to affect the behavior of the system. For simplicity, we will also assume it to be constant over species, $m_i=m$.
 
More precisely, the continuous variable $N_i(t)$ denotes the relative species abundance at time $t$ obtained by normalizing the population by the total number of individuals $N_\text{ref}$ that would be present in the absence of interaction. 
The underlying discrete birth-death process in population dynamics is then approximated by a continuous formalism embedding a multiplicative Gaussian demographic noise, whose covariance is inversely proportional to $N_\text{ref}$. 
The demographic stochasticity variable $\eta_i$ -- accounting for deaths, births, and other unpredictable events\footnote{The noise should be interpreted through Ito's convention since we want it to depend on the abundance at previous times. In this way, we correctly find that if a species is extinct it remains so at any later time (in the absence of immigration).} -- is then modelled by a white noise with zero mean and variance $\langle \eta_i(t)\eta_j(t')\rangle=2TN_i(t)\delta_{ij}\delta(t-t')$, with $T\propto 1/N_\text{ref}$ \footnote{The larger the global population, the smaller the strength, $T$, of the demographic noise \cite{altieri2021a}}.

$V_i(N_i)$ appearing in Eq. (\ref{dynamicaleq}) represents a single-species \emph{potential}, which in the logistic growth Lotka-Volterra model is quadratic in $N_i$:
\begin{equation}
	V_i(N_i)=-\rho_i\left(k_iN_i-\frac{N_i^2}{2}\right)
\end{equation}
The dynamics controlled by this potential has two fixed points, respectively in zero (unstable) and in the carrying capacity, $k_i$ (stable). The timescale at which the stable equilibrium is reached is fixed by the parameter $\rho_i$.
In the following we shall take $\rho_i=1$, and $k_i=k$ constant over species. 
Our results are easily generalizable to the case in which $\rho_i$ and $k_i$ are normally distributed \cite{biroli2018}. We will come back to this point and to a possible variation of this scenario.

Following pioneering works by R. May \cite{May1972}, the system parameters are usually sampled from random distributions, which make tools from statistical mechanics of disordered systems ideally suited. 
Nonetheless, at variance with Ref. \cite{May1972}, where the random sampling is performed at equilibrium, we consider here a pool of species with reciprocal random interactions \cite{MacArthur2016, Rieger1989, bunin2017, Galla2018} and we start from the hypothesis that the final equilibrium composition is a result of the dynamics. Random interactions are thus encoded in the matrix $\alpha_{ij}$ whose elements are drawn from a Gaussian distribution and assumed to be symmetric\footnote{The introduction of random asymmetric interactions would correspond to non-conservative forces in the dynamics and therefore give rise either to limit cycles or chaotic behavior.}, $\alpha_{ij}=\alpha_{ji}$. Even though it can appear to be a quite strong assumption, the symmetric case turns out to reasonably well describe communities in which the interactions stem from competition for resources or mutualistic behavior \cite{Blanchard2015, Carr2019}. 
In the following, we will focus on the competitive case considering positive values of the parameter denoting the mean interaction; negative values would instead favor cooperation.

In other words, Eq. (\ref{dynamicaleq}) represents a set of generalized Langevin equations for which, in the case of purely symmetric couplings, it can be shown that the corresponding Fokker-Planck equation admits a stationary probability distribution with associated temperature $T$. We will therefore denote in the following the zero temperature case as the process of averaging the abundances over an infinite population.
Both cases of zero and finite temperature have been recently investigated using disordered system techniques \cite{bunin2017, biroli2018, altieri2021a}.
The phase diagram at zero temperature was first obtained within the cavity method in \cite{bunin2017}, in which an additional parameter allowing also for asymmetric interactions was considered.
Novel results, related to the introduction of a finite demographic noise and immigration, were then discussed in \cite{altieri2021a}.

When the interactions are not too widely distributed, a single equilibrium, that is always reached no matter the assembly history, is found. Conversely, at high enough heterogeneity of the interactions, upon decreasing the amplitude of the demographic noise the system undergoes a phase transition characterized by the emergence of multiple equilibria. 
Decreasing further the noise, a second phase transition can take place below which all states are marginally stable and hierarchically organized \cite{altieri2021a}, hence underscoring several analogies with low-temperature glassy phases. 

\subsection{Meaningful scaling with $S$}
Two different choices have been considered in the literature for the scaling of the mean and the variance of the distribution of the interactions $\alpha_{ij}$ with the number of species.
In references \cite{bunin2017, biroli2018, altieri2021a}, the $\alpha_{ij}$ are extracted independently from a distribution with $\text{mean}[\alpha_{ij}]=\mu/S$ and $\text{var}[\alpha_{ij}]= \sigma^2/S$, where $\mu$ and $\sigma$ are taken to be finite in the large $S$ limit.
With this choice, the total effect on one species of the interaction with all others can remain finite even with an infinite number of non-extinct species, and we can find an analytical solution to the problem.
In references \cite{Kessler2015, fried2016, hu2021} the interactions are assumed not to scale with the number of species, as it is natural to do if we assembly an ecosystem by artificially adding species to the community \cite{hu2021}. In \cite{Kessler2015}, the introduction of asymmetric interactions can lead to a chaotic phase in which there is a sharp separation between many rare species and a few abundant ones with sudden turnovers between the two.
In \cite{Baron2019} it was shown that in this chaotic phase the abundances distribution satisfies a power-law distribution, in agreement to what was observed in marine micro-organisms communities \cite{SerGiacomi2018} and long-term experiments with a complex food web composition \cite{Beninca2008}.
The main motivation of our work is to bridge the gap between these two different approaches.

In the case of symmetric interactions, we cannot observe chaotic dynamics but, as a first step, we can study the mapping of one scaling of the interactions into the other and try to clarify still unanswered questions.
If mean and variance scale as the inverse of the number of species, we can recover $\alpha_{ij}\sim 1$ by taking $\mu\propto S$, $\sigma\propto \sqrt{S}$.
We will thus explore the limit in which $\mu$ is large and consider both finite $\sigma$ and $\sigma\propto\sqrt{\mu}$.
We will see that, at least at zero temperature, taking $\mu\to\infty$, $\sigma\propto \sqrt{\mu}$ is indeed equivalent to considering $\alpha_{ij}\sim 1$.

We will determine which phase is reached in the different limits and characterize the average behavior of the abundance and its probability distribution. 

\section{A thermodynamic formulation}
\label{sec:literatureresults}

Considering the symmetric interaction case allows us to write a quasi-equilibrium stationary probability distribution \cite {biroli2018} with Hamiltonian
\begin{equation}
	H=\sum_i V_i(N_i)+\sum_{i<j}\alpha_{ij}N_iN_j+(T-m)\sum_i \ln(N_i)
\end{equation}
which reminds a spin-glass structure where the (continuous variables) abundances play the role of the spins.

Note also that to guarantee the probability distribution to be integrable at small $N_i$, we need $m>0$.
This reflects the fact that, in the absence of immigration, $N=0$ would be an absorbing boundary condition, and at any finite value of the demographic noise any species would sooner or later become extinct. 
Alternatively, we can take the limit of small immigration after the $T\to0$ limit.

In the thermodynamic limit, as $S \rightarrow \infty$, physical quantities do not depend on the realization of the disorder $\alpha_{ij}$.
In particular, the free energy, which contains information about macroscopic equilibrium properties of the system, is self-averaging \footnote{For a self-averaging quantity typical realizations will have a free energy equal to its average over the disorder.}.
This can be computed through the replica method \cite{Mezard1986}, a standard tool in disordered systems. 
In this way, we can fully characterize the phase diagram both at zero and at finite temperature.
The main passages of the solution according to the replica formalism, adapted from references \cite{biroli2018, altieri2021a}, are presented in Appendix \ref{app:solution}. 
To have an overall picture, in the following we shall perform in-depth analysis in the case of strongly competitive interactions, which is still poorly understood. We shall mostly focus on the single equilibrium phase, firstly because we will see that this is the only relevant phase in the strong interaction limit at finite temperature, and secondly because at zero temperature we have a direct transition to the critical multiple-equilibria regime. Here the analytical computations become extremely involved and beyond the scope of this work.

\subsection{The single equilibrium phase}

Once the thermodynamic problem is well-defined, we can resort to the replica method to obtain an effective Hamiltonian for a single species abundance. In the single equilibrium phase the effective Hamiltonian reads (more details in Appendix \ref{app:solution}):
\begin{equation}
\begin{split}
	H_{eff} = \left[ 1-\sigma^2 \beta(q_d-q_0) \right]N^2/2 + \\+(\mu h-k-z \sqrt{q_0}\sigma)N + (T-m)\ln N
	\end{split}
	\label{eff_H}
\end{equation}
where $\beta=1/T$ denotes the inverse temperature.
According to Eq. (\ref{eff_H}) the single species abundance is subject to an effective Hamiltonian that, in addition to the single-species potential, contains some extra terms due to the mean field interactions with the rest of the system.
One of these terms is proportional to a Gaussian fluctuating field $z$, that accounts for the randomness of the interactions.
The parameters $h$, $q_d$ and $q_0$ are self-consistently fixed to be first and second moments of the distribution of $N$:
\begin{equation}
	\label{eq:selfcons1}h = \overline{\langle N \rangle} =
		\int \mathcal{D}z \left(\frac{\int_0^\infty dN e^{-\beta H_{eff}(q_0, q_d, h, z)}N}{\int_0^\infty dN e^{-\beta H_{eff}(q_0, q_d, h, z)}}\right)
		\end{equation}
		\begin{equation}
		\label{eq:selfcons2}
	q_d  = \overline{\langle N^2 \rangle}=\int \mathcal{D}z \left(\frac{\int_0^\infty dN e^{-\beta H_{eff}(q_0, q_d, h, z)}N^2}{\int_0^\infty dN e^{-\beta H_{eff}(q_0, q_d, h, z)}}\right) 
	\end{equation}
	\begin{equation}
	\label{eq:selfcons3}
	q_0 = \overline{\langle N \rangle^2}=\int \mathcal{D}z \left(\frac{\int_0^\infty dN e^{-\beta H_{eff}(q_0, q_d, h, z)}N}{\int_0^\infty dN e^{-\beta H_{eff}(q_0, q_d, h, z)}}\right)^2 
\end{equation}
where we have used the calligraphic notation for the Gaussian integral in $z$, i.e. $\int\mathcal{D}z\equiv\int_{-\infty}^{\infty}\frac{dz}{\sqrt{2\pi}}e^{-z^2/2}$.
The brackets indicate the average over the Boltzmann distribution for $N$ with the effective Hamiltonian $H_{eff}$, while the overbar stands for the average over the disorder, i.e. over the Gaussian variable $z$.
These averages coincide with the thermal and disorder average for a single species abundance.

In the $T\to 0$, $m\to0$ limit the thermal averages are dominated by the value $N^*(z)$, which minimizes the energy:
\begin{equation}
	\label{eq:Nstar}
	N^*(z)=\max\left\{0, \frac{\sigma\sqrt{q_0}}{1-\sigma^2\Delta q}(z+\Delta)\right\} \ ,
\end{equation}
with
\begin{equation}
\Delta=\frac{k-\mu h}{\sqrt{q_0}\sigma} \ .
\end{equation}
This greatly simplifies the analytical and numerical treatment of the equations, since only one average has to be performed. Equations in the $T \rightarrow 0$ limit were first obtained in \cite{biroli2018}. We recap here the main points. 

At small temperature $q_d-q_0\propto T$, so that the relevant parameter to study is $\Delta q= \beta(q_d-q_0)=\beta\overline{\langle N^2\rangle-\langle N\rangle ^2}$. 
The set of self-consistent Eqs. (\ref{eq:selfcons1})-(\ref{eq:selfcons2})-(\ref{eq:selfcons3}) can thus be rewritten as\footnote{The equation for $\Delta q$ is obtained computing separately the variance of $N$ in the case of extinction ($N^*=0$, $\beta(\langle N^2\rangle-\langle N\rangle ^2)\to0$) and of survival ($N^*>0$, $\beta(\langle N^2\rangle-\langle N\rangle ^2)\to1/H^{''}_{RS}$).}:
\begin{align}
	\begin{aligned}
		\label{eq:selfconsT0}
	h=&\overline{N^*(z)}=\frac{\sqrt{q_0}\sigma}{1-\sigma^2\Delta q}w_1(\Delta)\\
	q_0=&\overline{N^*(z)^2}=\frac{q_0\sigma^2}{(1-\sigma^2\Delta q)^2}w_2(\Delta)\\
	\Delta q=& \overline{\frac{\theta (N^*(z))}{H_{RS}^{''}(N^*(z))}}=\frac{1}{1-\sigma^2\Delta q}w_0(\Delta)
	\end{aligned}
\end{align}
where the $w_i$ are given by
\begin{align}
	w_i(\Delta)&=\int_{-\Delta}^\infty \frac{dz}{\sqrt{2 \pi}}e^{-z^2/2}(z+\Delta)^i
\end{align}
Such self-consistent equations can be numerically solved by iteration: we fix an initial guess for $h$, $q_0$ and $\Delta q$ and we compute the updated values by applying the formulas in (\ref{eq:selfconsT0}). To reach a faster convergence, some damping is added to the iteration protocol. The algorithm keeps going until a fixed point is reached, this will be a solution of the self-consistent equations. 

The abundance of one species $N^*$ depends linearly on the random value of $z$, unless this is smaller than the threshold value $-\Delta$, below which we have the species extinction, with $N^*=0$. 

The species abundance probability distribution is thus a Gaussian truncated in zero, plus a $\delta$ function in zero with weight equal to the fraction of extinct species (for more general results see Fig. \ref{fig:pdfT0} later).
The weight of the Gaussian part corresponds to the diversity $\phi$, i.e. the fraction of non-extinct species, given by 
\begin{equation}
	\phi=\overline{\theta(N^*)}=w_0(\Delta)
\end{equation}
which can be exactly computed in terms of error functions.

%The probability distribution of the mean abundance in the two temperature regimes is further discussed in Section \ref{app:crossover}.

\subsubsection{Stability analysis}

To check the stability of the single equilibrium solution we can compute the matrix of harmonic fluctuations of the replicated free energy close to this equilibrium point. The structure of its first vanishing eigenvalue -- defined in the replica jargon as \emph{replicon eigenvalue} -- is known \cite{Almeida1978}. In the single equilibrium phase it is given by \cite{biroli2018}:
\begin{equation}
	\lambda = (\beta \sigma)^2\left(1-(\beta \sigma)^2\overline{(\langle N^2 \rangle-\langle N \rangle^2)^2}\right) \ .
\end{equation}
When this stability eigenvalue touches zero, the replica symmetric solution becomes unstable, leading to a more complex landscape structure \cite{Mezard1986,DeDominicis2006}. The marginality condition, $\lambda=0$, is a key determinant for obtaining the transition line (this instability line will be shown in orange in Fig. \ref{fig:phasediagram} while presenting our results in detail).

Conversely, at zero temperature we would always obtain a diverging eigenvalue because of the multiplying $\beta^2$ factor. Since we are only interested in its sign, we rescale it as $\widetilde{\lambda}=\lambda/(\beta^2\sigma^2)=1-(\beta \sigma)^2\overline{(\langle N^2 \rangle-\langle N \rangle^2)^2}$.
Performing a computation similar to the one for $\Delta q$, $\widetilde{\lambda}$ may be computed analytically, leading to:
\begin{equation}
	\widetilde\lambda = 1-\sigma^2\overline{\frac{\theta (N^*(z))}{H_{RS}^{''}(N^*(z))^2}}=1-\sigma^2\frac{w_0(\Delta)}{(1-\sigma^2\Delta q)^2} \ .
\end{equation}

\section{Results: zero-temperature scenario}
\label{sec:zeroT}

In the following Sections, we aim to analyze previously unexplored regimes completing the phase diagrams and investigating systematically the dependence of the control parameters on high-$\mu$ values and thermal fluctuations.

\subsection{Finite number of species}

To explicitly explore the case in which $\mu\propto S$, $\sigma^2\propto S$, it is useful to consider first a finite number of species $S$.

The self-consistent expressions for $h$, $q_0$, and $q_d$ in Eqs. (\ref{eq:selfconsT0}) were obtained as the stationary conditions from a saddle-point approximation in the limit $S\to\infty$.
We can perform the same procedure by taking the limit $\beta\to \infty$ instead of $S\to \infty$ (see Appendix \ref{app:finitesize}).
Since $S$ is a finite quantity, we can now simply fix $\mu=\hat\mu S$, $\sigma^2=\hat{\sigma}^2S$.
The only relevant dependence on $S$ is now through $\mu$ and $\sigma$, therefore taking the $S\to \infty$ limit is equivalent to taking the limit $\mu\to \infty$, $\sigma\to \infty$, $\sigma\propto\sqrt{\mu}$. We conclude that, at least at zero temperature, this naive approach is exact.

%confirming as final result Eqs. (\ref{eq:selfconsT0}) above.

\subsection{Analytical solution in the thermodynamic limit}

In the following we generally work in the thermodynamic limit, for $S \rightarrow \infty$, and consider various scalings for $\mu$ and $\sigma$, knowing that at zero temperature this mapping is exact.
The phase transition at $T=0$ occurs at a constant value of $\sigma_c=\frac{1}{\sqrt{2}}$ for any value of $\mu$ \cite{bunin2017, biroli2018}.

% To check whether the replica symmetric ansatz is correct, we need to verify that the replicon is positive. 

% We numerically see that $\widetilde \lambda>0$ for $\sigma<\sigma_c=\frac{1}{\sqrt{2}}$ (Figure \ref{fig:repliconT0}), this is in agreement with the analytical results of references \cite{bunin2017, biroli2018}.
% Therefore we have that at any value of $\mu$ the solution is replica symmetric for $\sigma<\sigma_c$, while replica symmetry is broken above this value.

% Even though the fact that the critical value of $\sigma$ is unchanged in the $\mu\to\infty$ limit was expected from the results in the literature, it is still interesting to notice that the fluctuations of the interactions play such a critical role even when they are much smaller than the average.

% Looking also at the $\sigma\to \infty$ limit, as we should do to approach the case in which $\alpha_{ij}\sim 1$, would always result in a replica symmetry broken phase or in the unbounded growth phase, depending on how the two limits are taken.
% \begin{figure}[htbp]
% \centering
% \includegraphics[width=0.5\textwidth]{hq0sigmacloglog}
% 	\caption{\small{Numerical (dots) and analytic (line) solution for $h$ and $q_0$ at $\sigma=\sigma_c$. {\color{red}Remove?}
% }}
% \label{fig:ansigmac}
% \end{figure}
Along the transition line, we find the following expressions
\begin{align}
	h=\frac{k}{\mu} &&
	q_0=\pi\frac{k^2}{\mu^2} &&
	\Delta q=1 \ ,
\end{align}
which solve the self-consistent equations above.
As it is expected, we obtain that $\widetilde\lambda = 0$. The fraction of non extinct species $\phi$ is equal to $1/2$ at the transition, in agreement with one of the marginality conditions found in \cite{biroli2018}:
\begin{align}
	\phi\sigma^2=\frac{1}{4} && 	\sigma^2\Delta q=\frac{1}{2}
\end{align}
The first condition represents May's bound for the limit of stability \cite{May1972}: at a stable equilibrium the variance of the interactions, $\sigma^2/S$, should be smaller than the inverse of the number of surviving species, $S\phi$. The factor $1/4$ is because we are considering symmetric interactions so that the variance is effectively smaller. May's bound is therefore saturated at the transition, this property also holds in the entire multiple equilibria phase.
Also the second marginality condition is fulfilled by our solution.

We can then expand the equations close to the transition point in powers of $d\sigma=\sigma-\sigma_c$; solving them to the first order we obtain:
\begin{align}
	\begin{aligned}
	\Delta=&\sqrt{\frac{\pi}{2}}d\sigma\\
	h=&\frac{k}{\mu}\left(1+\frac{\sqrt{2}}{\mu} d\sigma\right)\\
	q_0=&\pi\frac{k^2}{\mu^2}\left(1+2\sqrt{2}(\pi/\mu-2+\pi)d\sigma\right)\\
	\Delta q=&1-\sqrt{2}d\sigma\\
	\widetilde\lambda =& 2\sqrt{2}d\sigma
	\end{aligned}
\end{align}
The results are in good agreement with the numerical solutions at linear order around $\sigma_c$. We refer the interested reader to Appendix \ref{numerics}.
The expansion confirms that to the leading order the average abundance at zero temperature is inversely proportional to $\mu$, also away from the transition line.

We also obtain that the diversity $\phi$ is independent of $\mu$.
Therefore, the vanishing trend of the average value of the abundances at large $\mu$ is not due to the extinction of a large fraction of species, but to a general decrease in the typical value of the abundances. 
Indeed, the truncated Gaussian describing the non-zero part of the distribution is centered around a quantity proportional to $k-\mu h$ (that is exactly zero at the transition and tends to zero at large $\mu$ for any $\sigma$), and has a width proportional to $\sqrt{q_0}$, that goes to zero at large $\mu$ (see also Figure \ref{fig:pdfT0}).
The exact cancellation at the transition between the carrying capacity $k$ and the mean effect of other species $\mu h$ is the reason why a finite $\sigma$ can play a crucial role even in the limit of large mean interaction.

Taking $\sigma \propto \sqrt{\mu}$ to recover the scaling $\alpha_{ij}\sim 1$ we always end up in the critical phase or in an unbounded growth region. Indeed, increasing further the heterogeneity of the interactions results in a pathological unbounded growth of the abundances, due to the emergence of a subset of species with cooperative interactions strong enough to override the single-species saturation. This effect can be cured by adding a stronger divergence at large $N$ in the single species potential, but this possibility goes beyond the aim of this paper. For more details about a cubic potential see \cite{Altieri2021}, which accounts for mutualistic interactions under some specific assumptions.

\section{Results: Finite temperature}
\label{sec:finiteT}

Going back to finite temperature, in addition to the disorder average we have to consider the thermal one, with the effective Hamiltonian of equation \ref{eff_H}. 
This contains a quadratic and a logarithmic term; a quadratic Hamiltonian would allow us to explicitly compute the thermal averages, and to make some analytical progress on the solution.
Therefore, to eliminate the logarithmic contribution we can use two different strategies:
\begin{itemize}
    \item {We can set $m=T$, but this case is a bit pathological since the slightest change in $m$ would result in a diverging or zero value of the probability distribution in $N=0$. Furthermore, to really understand the role of temperature we would like to be able to vary it independently from the immigration rate.}
\item{Alternatively, the term also drops if we take the $T\to0$, $m\to0$ limit as we did in the previous section. Neglecting the logarithmic term in the Hamiltonian but keeping a finite $\beta$ allows us to artificially study the Hamiltonian that controls the system at zero temperature, but in the presence of fluctuations. This corresponds to effectively disentangling the effect of demographic noise on the potential and the temperature that appears in the Boltzmann weight.}
\end{itemize}
The case in which we neglect the logarithmic term is also of interest because it can be directly mapped into the replicator equations \cite{Diederich1989, Biscari1995}. The mapping between the Random Replicant Model and the Lotka-Volterra equations has been put back on the table recently and shown exactly in the thermodynamic limit \cite{altieri2021a}. 

In the following, we will neglect this logarithmic term, keeping in mind this double interpretation of what we are doing.
Then, in the single-equilibrium phase the quadratic Hamiltonian can be rewritten as 
\begin{equation}
	H_{eff}=\frac{d}{2}(N-N_0)^2
\end{equation}
where $N_0$ is linear in the Gaussian field $z$ and $d$ is a constant:
\begin{align}
\begin{aligned}
	d&=1-\sigma^2\beta(q_d-q_0)\\
	N_0&=-\sqrt{\beta}\frac{\mu h -k-z\sigma\sqrt{q_0}}{\sqrt{2d}} \ .
\end{aligned}
\end{align}
The factor $d$ controls the concavity of the Hamiltonian, so that we need it to be positive to have a stable system, while $N_0$ represents the minimum of the parabola. The positive constraint on the abundances implies that if $N_0<0$ the minimum of the energy will shift to $N=0$.

% \begin{figure}[htbp]
% \centering
% \includegraphics[width=0.5\textwidth]{hq0qdanloglog.pdf}
% 	\caption{\small{Numerical results of equations (\ref{eq:selfcons1}) (dots) and analytical results at dominant order in $1/\mu$ (continuous lines, equations (\ref{eq:hq0qdan})). $\sigma=1/5$, $\beta=1$.} {\color{red}Remove?}
% }
% \label{fig:qmu}
% \end{figure}

As it can be checked \emph{a posteriori}, we assume that $h$ goes to zero more slowly than $k/\mu$ in the large $\mu$ limit. $N_0$ then diverges to $- \infty$ allowing us to perform an expansion of the self-consistent equations and to obtain the asymptotic expressions:
\begin{align}
	\begin{aligned}
		\label{eq:hq0qdan}	
		h&\approx\frac{1}{\sqrt{\beta\mu}}\\
		q_0&\approx\frac{1}{\beta \mu}\\
		q_d&\approx\frac{2}{\beta \mu}
	\end{aligned}
\end{align}
The derivation, valid for $\mu\gg\beta$, is detailed in Appendix \ref{app:finitetemperatureexp}.
These results are in good agreement with the numerical solution at small $\beta$ and large $\mu$ (Fig. \ref{fig:hvarbeta}), as we specifically proved for this work.

The disorder, represented by the fluctuating field $z$, plays no role in this limit: the term containing $z$ is simply dropped in the expansion by noting that $\mu h -k\gg\sigma\sqrt{q_0}$. 
In other words, the constant terms in the effective Hamiltonian are so large that fluctuations become negligible.
Moreover, at variance with the zero temperature regime, here the scale of the populations is independent of the carrying capacity $k$: it is determined by the thermal fluctuations and the immigration, which turn out to be strictly linked.

We can also compute the stability eigenvalue (see Appendix \ref{app:replicontfinite} for more details), obtaining:
\begin{equation}
	\label{eq:repliconmugrande}
	\lambda = (\beta \sigma)^2\left(1-\frac{\sigma^2}{\mu^2}\right)>0
\end{equation}
We see that it is always positive in the limit of large $\mu$, indicating that the single equilibrium solution that we have studied so far is correct. Finding always a single equilibrium means that the disorder is irrelevant in this regime.

\subsection{Crossover and comparison between the two temperature regimes}
\label{app:crossover}

\begin{figure}[htb]
	\centering
	\begin{subfigure}[b]{0.235\textwidth}
		\centering
		\includegraphics[width=\textwidth]{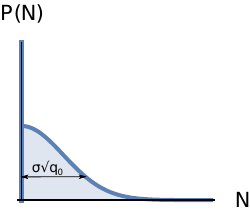}
		\caption{\small $T=0$}
		%\caption[]{\small Probability distribution of the abundance of a given species at zero temperature. It is composed by a $\delta$ function in zero with weight $1-\phi$ and a truncated Gaussian centered in $(k-\mu h)/(1-\sigma^2 \Delta q)$ and with width $\sigma \sqrt{q_0}/(1-\sigma^2\Delta q)$.}
		%{{\small Denominator, $\beta=1000$.}}    
	{\small {} }
	\label{fig:pdfT0}
\end{subfigure}
\begin{subfigure}[b]{0.235\textwidth}
		\centering 
		\includegraphics[width=\textwidth]{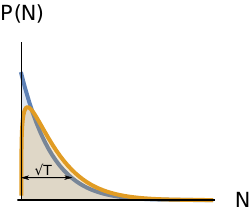}
		\caption{\small $T>0$}
		%\caption[]{\small Probability distribution of the abundance of a given species at finite temperature, for $\lambda=T$ (blue) and reintroducing a small positive logarithmic (orange).}%
		%{{\small Replicon eigenvalue, $\beta=1000$.}}    
		\label{fig:pdfTfinite}
\end{subfigure}
	\caption[]{\small Qualitative sketches of the probability distribution of the abundance of a given species at zero (a) and finite temperature (b). At finite temperature we show the  distribution for $m=T$ (blue) and reintroducing a small positive logarithmic term (orange).}%
	\label{fig:pdfs}
\end{figure}

It is interesting to compare the probability distributions of the abundances in the two regimes. 
In both cases, we have a truncated Gaussian (see Fig. \ref{fig:pdfs}), but they have different origins. 
At zero temperature the finite width of the distribution is due to the disorder, which entirely determines the value of the abundance. The distribution is composed of a truncated Gaussian -- centered close to zero and with width proportional to $\sqrt{q_0}\sigma$ -- and of a $\delta$ function in zero with weight equal to the truncated part of the Gaussian, $1-\phi$.
This distribution is often found in disordered systems with hard constraints (here represented by $N\geq0$); for example, it arises in resource competition ecosystems \cite{Tikhonov2017, Altieri2019}, in the computation of the storage capacity of a neural network \cite{Gardner1988}, and in the study of optimal equilibria in complex economies \cite{DeMartino2004}.

At finite temperature in the strong interaction limit, the effect of disorder is negligible, and the variability in the abundances is only due to thermal fluctuations. 
The distribution is the tail of a Gaussian centered in $N_0$ -- tending to $-\infty$ at large $\mu$ -- and of variance $1/(\beta d)$.
By lowering the temperature, the tail of the Gaussian becomes narrower and narrower and approaches a $\delta$ function. At the same time, the peak of the Gaussian moves to finite values, the $z$ dependence becomes relevant and causes a spread of the peak of the $\delta$ function. We recover then the zero-temperature distribution.
The probability distribution is the tail of a Gaussian only if we neglect the logarithmic term, therefore at exactly $m=T$. If we reintroduce a small positive logarithmic contribution the probability distribution is forced to pass through zero at $N=0$ (in orange in Figure \ref{fig:pdfTfinite}); on the other hand, with a negative logarithmic term we would have a divergence in $N=0$, integrable if $m>0$.

While at zero temperature a finite fraction of the species is extinct, at finite temperature they all have a positive, albeit small, abundance. 
This is a consequence of the fact that in this case we cannot take the limit $m\to 0$; the finite immigration results in finite populations. 

We can relax the assumption of constant carrying capacities by taking them to be normally distributed. Our results easily generalize to this case: we would still obtain Gaussian distributions, but with a corrected variance \cite{biroli2018}. We expect instead that considering a lognormal distribution for the carrying capacities, as suggested in \cite{grilli2020}, would strongly impact our predictions leading to a wider distribution also for the abundances.

We have seen that at zero and finite temperature we obtain two different asymptotic behaviors: at $T=0$ we have $h\sim k/\mu$, while at finite temperature $h\sim1/\sqrt{\beta\mu}$.
Therefore, we have also studied numerically the crossover between the two regimes by considering increasing values of $\beta$.

\begin{figure}[htbp]
\centering
\includegraphics[width=0.475\textwidth]{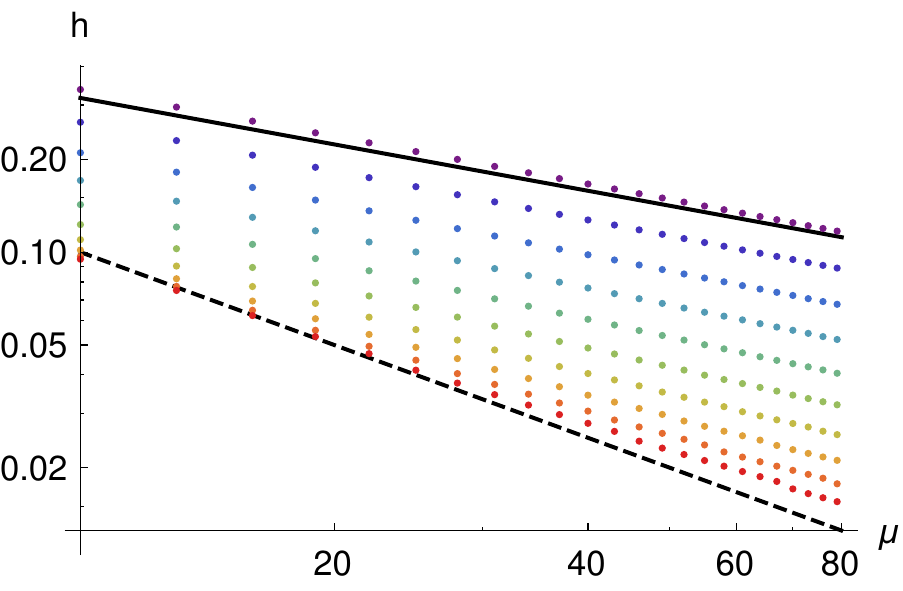}
	\caption{\small{Numerical results for the average abundance $h=\overline{\langle N\rangle}$ as a function of $\mu$ for geometrically increasing values of $\beta$, from $\beta_{min}=1$ (purple) to $\beta_{max}=200$ (red). The full line (shown for $\beta=\beta_{min}$) and the dashed line represent the asymptotic results for $\mu\gg\beta$ for $\mu\ll\beta$ respectively.}}
\label{fig:hvarbeta}
\end{figure}
%and to recover the zero temperature solution in the formalism we used for the finite temperature one.
%We can again consider the expression (\ref{eq:meanNz}) for the thermal average of the abundance at fixed $z$ and consider the limit $\beta\to\infty$. We find that $c$ diverges at finite $\mu$ to $+\infty$ or $-\infty$ depending on the value of $z$, so that we obtain as before:
%\begin{equation}
	%\label{eq:meanNz}
	%\langle N\rangle(z)\rightarrow
	%\begin{cases}
		%\frac{1}{\sqrt{2D}c}=\frac{1}{\beta}\frac{1}{\sigma\sqrt{q_0}(-z-\Delta)}\to 0 &\text{if $z<-\Delta$}\\
		%-\sqrt{\frac{2}{D}}c=\frac{\sigma\sqrt{q_0}}{1-\sigma^2\beta(q_d-q_0)}\left(z+\Delta\right) &\text{if $z>-\Delta$}
	%\end{cases}
%\end{equation}

%We thus recover the zero temperature solution (\ref{eq:Nstar}). 
As we can see in Fig. \ref{fig:hvarbeta}, we can continuously interpolate between the two solutions: when $\mu\gg\beta$ the numerical solution is in good agreement with the finite temperature asymptotic expansion (purple dots and continuous line), while for $\mu\ll\beta$ we approach the zero temperature case (red dots and dashed line). In between, we have a continuous crossover.

\subsection{Strongly heterogeneous interactions}

% \begin{figure}[htbp]
% \centering
% \includegraphics[width=0.6\textwidth]{hq0qdvarrsmall}
% 	\caption{\small{Numerical results for the values of $h$, $q_0$ and $q_d$ as a function of $\mu$, with $\sigma=r\sqrt{\mu}$ and $r$ from $0.01$ to $0.66$. The curves at different $r$ are perfectly superimposed, and in good agreement with the asymptotic results of eq. (\ref{eq:hq0qdan}).}}
% \label{fig:qmuvarr}
% \end{figure}

The solutions we have found in the limit of strong coupling at fixed $\sigma$ and $T$ do not depend on $\sigma$. Since we do not expect this to always be the case, we might wonder what is the limit of validity of these results. 

At zero temperature we have seen that taking $\sigma\propto\sqrt{\mu}\to \infty$ is equivalent to considering $\alpha_{ij}\sim 1$, we expect this behavior to hold also at small but finite temperature. 

The finite temperature expansion relies on the fact that
\begin{equation}
	\label{eq:c2}
	-N_0=\sqrt{\beta}\frac{\mu h -k-z\sigma\sqrt{q_0}}{\sqrt{2d}}
\end{equation}
is large for any value of $z$ that contributes significantly to the integral, i.e. for $z \sim O(1)$. This corresponds to requiring
\begin{align}
	\label{eq:b}\sqrt{\beta}\frac{\mu h -k}{\sqrt{2d}}\gg 1&&
	\frac{\mu h -k}{\sigma\sqrt{q_0}}\gg1
\end{align}
Assuming the scalings of Eq. (\ref{eq:hq0qdan}), these conditions are also met if we take $\sigma\propto\sqrt{\mu}$. For not too large values of the ratio $r=\sigma/\sqrt{\mu}$ we can numerically verify that the previously found solutions are indeed still correct.

Nevertheless, an important difference with the previously considered case is that now the second term in $d=1-\sigma^2\beta(q_d-q_0)$ does not go to zero at large $\mu$, but it tends to a constant:
\begin{align}
    \beta\sigma^2(q_d-q_0)\sim r^2\mu\beta\frac{1}{\mu \beta}=r^2
\end{align}
%{\color{red} Opzione 1}
%By increasing $r$, such a constant reaches one, which in turn implies a zero in the denominator of $N_0$ at finite $\mu$.
%What happens at this point depends strictly on $z$: for $z<-\Delta=\frac{\mu h-k}{\sigma\sqrt{q_0}}$ we have $N_0 \rightarrow -\infty$, whereas for $z>-\Delta$ we have $N_0\to \infty$.
%In this second case, the thermal average of the abundance diverges; since this happens in a finite portion of the disorder integration region, the average abundance diverges as well.

By increasing $r$, this constant reaches one, which in turn implies that the coefficient of the quadratic part of the confining potential goes to zero. We observe then a divergence in the average abundance and inevitably end up in the unbounded growth region.

Since the scalings we have obtained remain valid until we encounter this singularity, the stability eigenvalue is still given by Eq. (\ref{eq:repliconmugrande}), and in the scaling we are considering is again positive.
We can therefore conclude that at large $\mu$ and finite temperature we have a direct transition from the single equilibrium phase to the unbounded growth phase. 

	\begin{figure}[htb]
		\centering
				\begin{subfigure}[b]{0.21\textwidth}   
			\centering 
			\includegraphics[width=\textwidth]{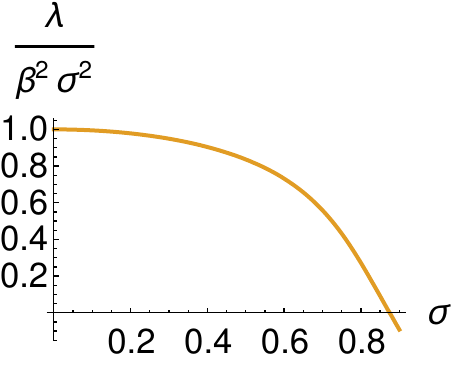}
			\caption[]{}%
			%{{\small Replicon eigenvalue, $\beta=1000$.}}    
			\label{fig:lambda0}
		\end{subfigure}
		%\hfill
		\begin{subfigure}[b]{0.21\textwidth}   
			\centering 
			\includegraphics[width=\textwidth]{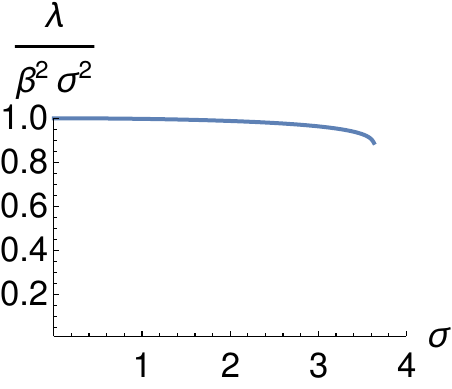}
			\caption[]{}%
			%{{\small Replicon eigenvalue, $\beta=1$.}}    
			\label{fig:lambdafinite}
		\end{subfigure}
		
		%\vskip\baselineskip

				\begin{subfigure}[b]{0.215\textwidth}
			\centering
			\includegraphics[width=\textwidth]{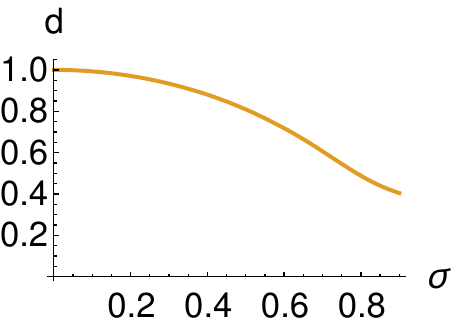}
			\caption[]{}
			%{{\small Denominator, $\beta=1000$.}}    
			\label{fig:denfinite}
		\end{subfigure}
		\begin{subfigure}[b]{0.215\textwidth}  
			\centering 
			\includegraphics[width=\textwidth]{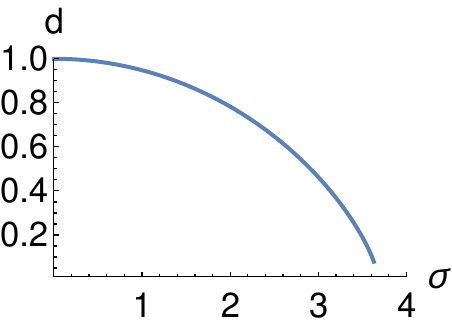}
			\caption[]{}%
			%{{\small Denominator, $\beta=1$.}}    
			\label{fig:den0}
		\end{subfigure}
				%\hfill
		\caption{{\small {Two different scenarios for the breakdown of the single equilibrium phase. At small temperature ($\beta=1000$, orange) we find a value of $\sigma$ for which the stability eigenvalue goes to zero, while the denominator $d$ is finite, hence ending up in the multiple equilibria phase. 
		At large temperature ($\beta=1$, blue) the denominator goes to zero before the stability eigenvalue: the unbounded growth phase is directly reached. To properly capture the behavior in the two-temperature regimes we use two different scales on the $x$-axis. In both cases, $\mu=20$.}}
		}
		\label{fig:repliconden}
	\end{figure}
	
		%\begin{figure}[htb]
	%	\centering
		%\begin{subfigure}[b]{0.21\textwidth}   	\centering 	\includegraphics[width=\textwidth]	{lambda20b1000fixed_axis}
			%\caption[]{}%
			%{{\small Replicon eigenvalue, $\beta=1$.}}    
		%	\label{fig:lambdafinite}
		%\end{subfigure}
			%	\begin{subfigure}[b]{0.21\textwidth}
		%	\centering
			%\includegraphics[width=\textwidth]{lambda20b1fixed_axis.pdf}
			%\caption[]{}
			%{{\small Denominator, $\beta=1000$.}}    
			%\label{fig:denfinite}
	%	\end{subfigure}

To further verify this claim we have studied the numerically integrated solutions of the self-consistent equations at fixed $\mu$ and $\beta$ upon varying $\sigma$.
In agreement with the previous analysis, we find two qualitatively distinct scenarios, that are shown in Figure \ref{fig:repliconden}. 
For $\beta\gg\mu$, where our asymptotic computation is not valid, increasing $\sigma$ the stability eigenvalue crosses 0 (Fig. \ref{fig:lambda0}) while $d$ still has a finite value (Fig. \ref{fig:denfinite}). 
This means that we are crossing to the multiple equilibria phase, in agreement with what is found in \cite{altieri2021a}. For $\beta\ll \mu$ we see instead that $d$ approaches zero (Fig. \ref{fig:den0}) while $\lambda$ has still a finite value (Fig. \ref{fig:lambdafinite}): we thus have a direct crossing to the unbounded growth phase.

\begin{figure}[htbp]
	\centering
	\begin{subfigure}[b]{0.38\textwidth}   
		\centering 
		\includegraphics[width=\textwidth]{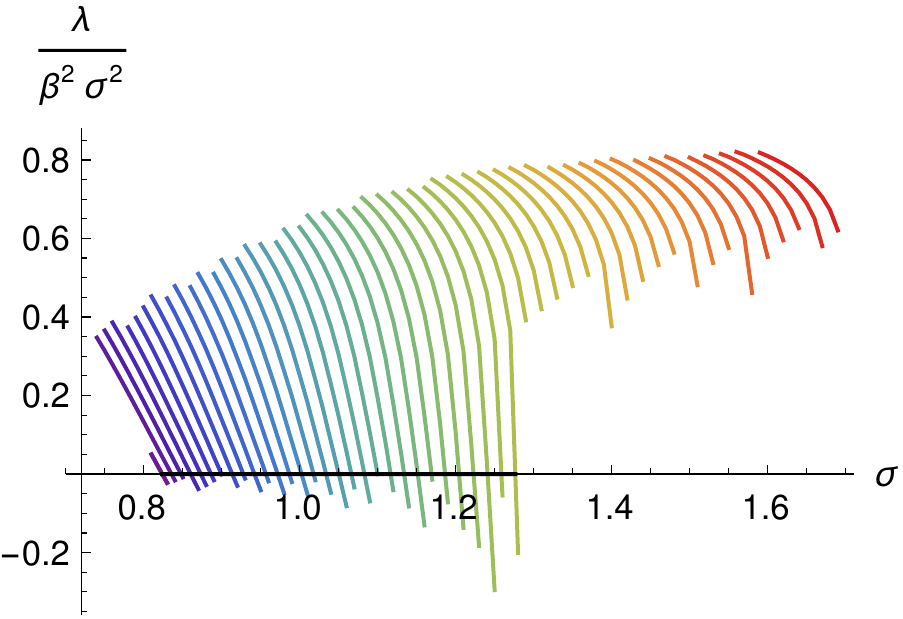}
		\caption[]%
		{\small Stability eigenvalue} 
		\label{fig:lambdavarmu}
	\end{subfigure}
	\hfill
	\begin{subfigure}[b]{0.38\textwidth}   
		\centering
		\includegraphics[width=\textwidth]{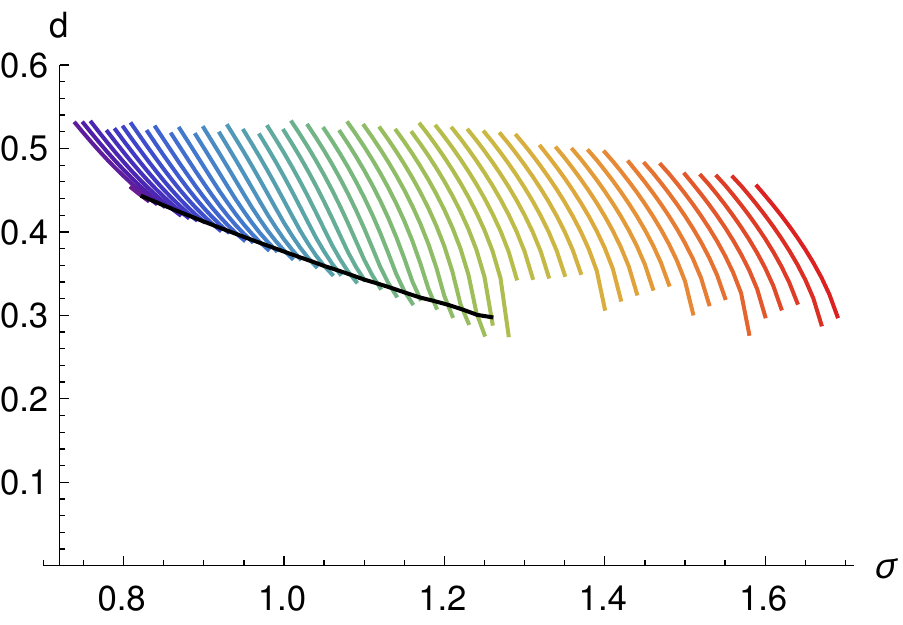}
		\caption[]%
		{{\small Denominator}}    
		\label{fig:denominatorvarmu}
	\end{subfigure}
	\caption{{\small Replicon eigenvalue and denominator of $N_0$ in eq. (\ref{eq:c2}) as a function of $\sigma$, for values of $\mu$ between 5 (purple) and 30 (red). The transition points -- shown in black -- correspond precisely to $\lambda=0$. Here $\beta=100$.}}
	\label{fig:mean and std of nets}
\end{figure}
We finally establish the phase diagram at fixed finite temperature, in the $\sigma$-$\mu$ plane: at fixed $\mu$ we progressively increase $\sigma$ until the stability eigenvalue reached zero obtaining a diverging result. In Fig. \ref{fig:lambdavarmu} we show the stability eigenvalue as a function of $\sigma$ for different values of $\mu$. This task is complicated by the fact that the numerical algorithm encounters overflow problems well before the denominator goes to zero, where a phase transition should occur (see Fig. \ref{fig:denominatorvarmu}). This is expected as the numerical protocol requires the computation of several exponentials and error functions. 

%\begin{figure}[]
	%\centering
	%\begin{subfigure}[b]{0.33\textwidth}   	\centering 
		%\includegraphics[width=\textwidth]{phasediagramtext}
	%	\caption[]%
	%	{\small $T>0$.} 
	%	\label{fig:phasediagram}
	%\end{subfigure}
	%\hfill
	%\begin{subfigure}[b]{0.37\textwidth}   
		%\centering 
		%\includegraphics[width=\textwidth]{phasediagramT0Diversitytext.pdf}
		%\caption[]%
		%{{\small $T=0$.}}    
	%	\label{fig:phasediagramT0}
	%\end{subfigure}
	%\caption[
	%{{\small Sketches of the phase diagram at finite and at 0 temperature. The bottom area is the single equilibrium phase, between the two lines we have the multiple equilibria phase, on top the unbounded growth phase.}}
	%\label{fig:phasediagramsT0finite}]
%\end{figure}

\begin{figure}[t]
	\centering
	\begin{subfigure}[b]{0.46\textwidth}   
		\centering 
		\includegraphics[scale=0.76]{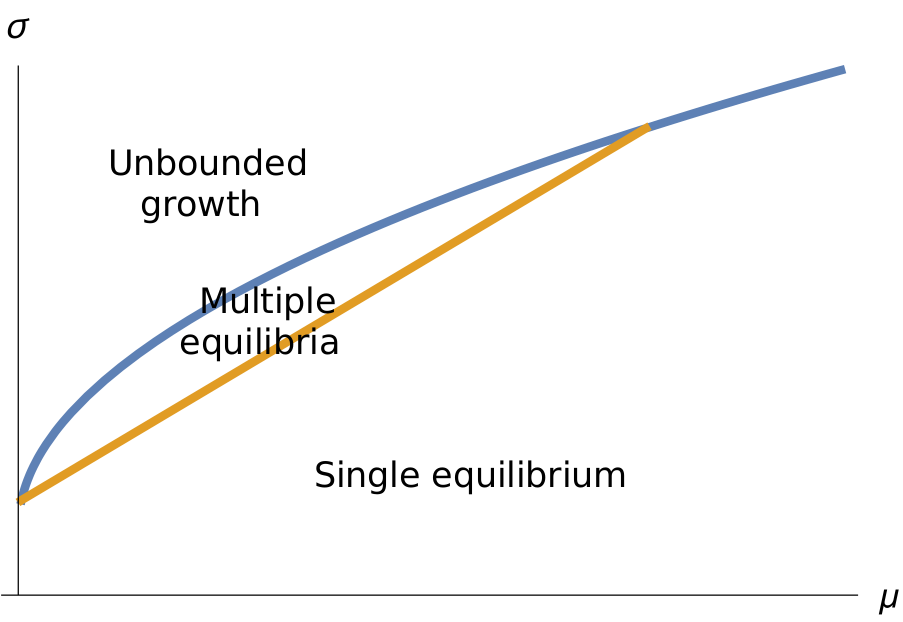}
		\caption[]%
		{\small $T>0$} 
		\label{fig:phasediagramT}
	\end{subfigure}
       \hfill
	\begin{subfigure}[b]{0.39\textwidth}   
		\centering
		\includegraphics[scale=0.695]{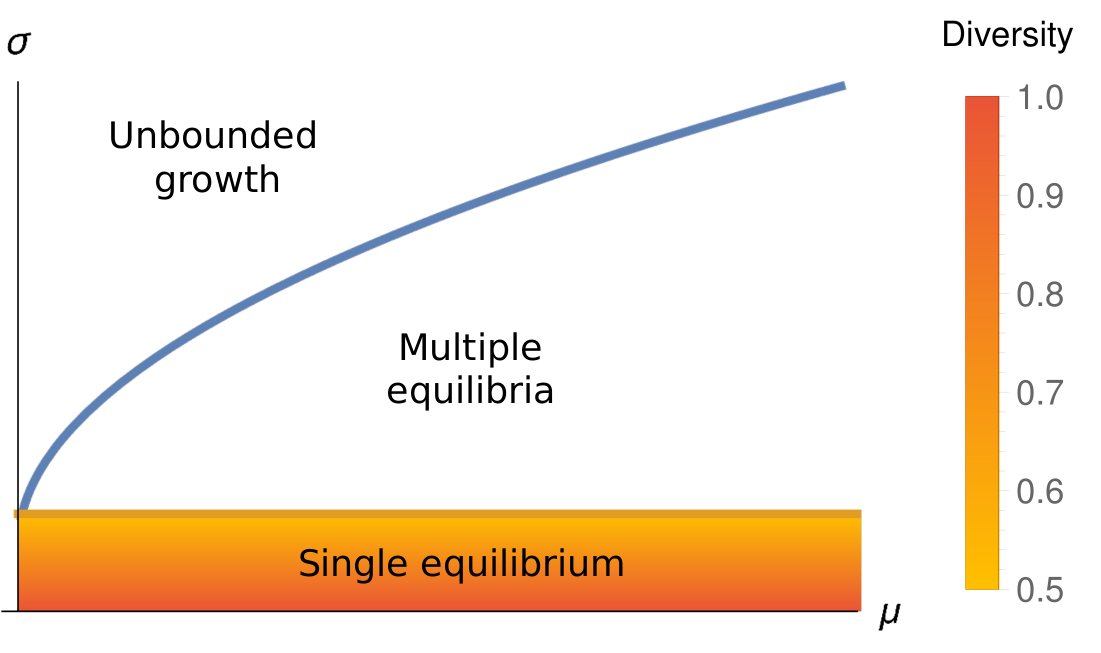}
		\caption[]%
		{{\small $T=0$}}    
		\label{fig:phasediagramT0}
	\end{subfigure}
\caption{\small{Qualitative phase diagrams at finite and zero temperature, respectively. The lowest part corresponds to the single equilibrium phase, whereas between the orange line and the blue one a multiple equilibria phase occurs. Above the blue line, the unbounded growth region is characterized by a divergence of the species abundances. In the $T \rightarrow 0$ limit, the diversity is also plotted in the single equilibrium phase.}}
	\label{fig:phasediagram}
\end{figure}
Nevertheless, at small enough values of $\mu$ we can reconstruct the transition curve, which appears to be a straight line in the $\sigma$-$\mu$ plane. 
This is in qualitative agreement with the asymptotic expression for the stability eigenvalue (Eq. \ref{eq:repliconmugrande}), that predicts a phase transition for $\sigma=\mu$. This line is never observed because at large $\mu$, where the asymptotic expression is valid, the divergence sets in before the transition.

Considering a square root behavior for the transition point to the unbounded growth region, which is what we expect at large $\mu$, we can give a qualitative representation of the phase diagrams, as shown in Figure \ref{fig:phasediagram}. In the bottom part we confirm the phase diagram obtained previously in the zero-temperature limit \cite{bunin2017} also for strong competition, \emph{i.e.} as $\mu \rightarrow \infty$. Note that at zero temperature the transition from the single equilibrium to the multiple equilibria phase occurs at a fixed value of $\sigma_c=1/\sqrt 2$, independent of $\mu$, while at finite temperature $\sigma_c$ grows linearly with $\mu$. Therefore it has to cross the transition line to the unbounded growth region, since this grows more slowly than linearly.
We can thus conclude that at finite temperature and strong interactions there is no multiple equilibria phase.

To estimate the order of magnitude of the value of $\mu$ at which the multiple equilibria phase disappears, we can consider the value of $d$ at the transition point, extracted from Figure \ref{fig:denominatorvarmu}. This parameter appears to decrease linearly as a function of $\mu$.
We expect the intersection between the transition line and the divergence line to occur when the value of $d$ at the transition reaches zero, we can thus get an estimate of the corresponding value of $\mu$ from a linear extrapolation, obtaining $\mu=39.1$.
This analysis must be taken as only qualitative since there is no guarantee that the linear behavior is maintained close to divergence. 

We also tried to compute the interception between the linear extrapolation of the critical value $\sigma_c$ and the square root curve $\sigma=\sqrt{\mu}$ that we expect to describe the transition to the unbounded growth phase at large $\mu$. We obtained a much larger value than with the previous method ($\mu=614$), confirming that the analysis should only be considered qualitative, because of the practical issues of approaching the divergences and the deviations from the asymptotic behavior at small $\mu$. 
In conclusion, taking $\sigma\propto\sqrt{\mu}\to\infty$ to recover $\alpha_{ij}\sim 1$ has dramatically different outcomes in the zero and finite temperature cases: at zero temperature we end up in the multiple equilibria phase, whereas at finite temperature we always remain in the single equilibrium phase. 

It is not easy to conclude what would actually happen for a finite system.
On the one hand, we have seen that the temperature is related to the inverse of the size of the system so that a finite system will always have a finite temperature, and therefore at large $\mu$ it would end up in the single equilibrium phase. 
On the other hand, we have seen that at zero temperature we can explicitly take $\mu\propto S$, so that a finite number of species would result in a finite $\mu$ associated with a possible multiple equilibria phase.
The outcome would therefore depend on how the size of the populations compares with the number of species, remembering that the crossover between the two asymptotic solutions is controlled by the ratio $\mu/\beta$.

Drawing an immediate comparison between our results and the numerical analysis performed in \cite{Kessler2015} is still difficult because the introduction of non-symmetric interactions leads also to chaotic behavior. Moreover, since the study in \cite{Kessler2015} relies on a relatively small number of species ($S \sim 20$), claiming to end up in any of the asymptotic regimes is not foregone.
We nevertheless expect that in the zero-temperature limit the asymmetry may be functional in transforming the multiple equilibria of the Hamiltonian into chaotic trajectories. Upon increasing the number of species the chaotic phase should progressively invade the entire phase diagram, in agreement with recent numerical and experimental results \cite{hu2021}.
Considering also demographic noise would contribute to adding stochastic fluctuations to the trajectories, hence pushing the transition to chaos and possibly merging the chaotic attractors into one \cite{altieri2021a}. Whether
the resulting chaotic phase is characterized by one or multiple attractors would depend on the relative strength of the demographic noise and its interplay with interaction heterogeneity\footnote{We expect the zero-demographic noise scenario to be relevant also in the case of TARA expedition data on planktonic ecosystems, as characterized by an enormous number of individuals \cite{SerGiacomi2018}, then allowing for chaotic regimes if asymmetry in the interactions is also taken into account.}.

\section{Conclusions and outlook}
\label{sec:conclusions}

In this work, we explored the limit of symmetric strongly competitive interactions in the random Lotka-Volterra model.
We showed that at zero temperature the strong competition limit (in which the distribution of the interactions does not scale with the system size) is equivalent to taking  $\sigma\propto\sqrt{\mu}\to\infty$. We therefore studied the limits in which the mean interaction $\mu\to\infty$, the heterogeneity parameter $\sigma\sim 1$, and $\sigma \propto \sqrt{\mu}$, at zero and finite temperature.
In both cases, the average abundance goes to zero upon augmenting the average interaction, but according to two different asymptotic behaviors. At zero temperature, the average abundance decreases as the inverse of the mean interaction, so that the average effect of the ecosystem on one species remains constant. 
%The most probable value of the abundance of non-extinct species approaches zero, thanks to an exact cancellation between the carrying capacity $k$ and the average interaction with other species $\mu \overline{\langle N\rangle}$; this allows the finite variance of the interactions to play a critical role.
The transition to the multiple equilibria phase is maintained and it occurs at the same value of the interaction heterogeneity as predicted for this model in the weakly interacting scenario. 
Therefore, considering a large heterogeneity to recover the strong competition limit would always lead the system to the multiple equilibrium phase. 
Interestingly, the decrease of the average abundance at fixed heterogeneity is not due to a loss of diversity, but to a general decrease of the typical abundance.
We then computed an expansion around the transition line, that is in good agreement with our numerical analysis at linear order.
At finite temperature, we performed an asymptotic expansion in the $\mu\to\infty$ limit, and obtained that the average abundance decays as $1/\sqrt{\beta\mu}$, in remarkable agreement with the numerical results. 
In the finite temperature case, we had to include a finite immigration rate to compensate for stochasticity-induced extinctions. This leads to larger populations, that result in a diverging average interaction, so that the finite fluctuations we were considering become irrelevant.
Consequently, we found that in this limit there is no phase transition at finite $\sigma$: the system appears to be always in the single equilibrium phase. 
Furthermore, by studying the crossover between the two regimes we found that in both cases the species abundance distribution decays as a Gaussian, showing that a key ingredient to recover power-law distributions in the abundances is the introduction of a full class of interactions, including unilateral ones.
An in-depth investigation of the asymmetric case -- that can be studied either perturbatively or by resorting to the cavity method for arbitrary values of the covariance between $\alpha_{ij}$ and $\alpha_{ji}$ -- is left for future research.
Alternatively, one can also complexify the model by including both demographic and environmental noise, regulated by two different scalings as a function of population size, which would allow us to identify regimes governed by power laws.

Finally, we considered the case in which $\sigma$ scales as $\sqrt{\mu}$, and recovered the previous solution for small enough values of the ratio $\sigma/\sqrt{\mu}$.
At large $\mu$, for some critical value of this ratio, still of $O(1)$, there is a direct phase transition to the unbounded growth phase.
We then claim that at finite temperature and large value of the average interaction no multiple equilibria phase shows up.
Having a more direct comparison with real-life ecosystems especially in terms of a multistability regime as pointed out in our theoretical phase diagram, appears to be a timely problem to look at. Dilution experiments in which tuning nutrient supply rates and environmental parameters are gaining momentum thanks to increasingly sophisticated techniques able to reproduce synthetic microbial communities \cite{hu2021}. Modifying the availability of nutrients -- which basically affects the strength of the interactions -- and the number of species can trigger abrupt changes in the community composition: these are signaled either by uncontrollable multiplication or by the emergence of several stable compositions if competing species have different stoichiometries with respect to the same essential nutrients \cite{Dubinkina2019}.

Although our analysis is restricted here to strong interactions, recent studies tend to support distributions of interaction strengths that are highly skewed -- with few strong and many weak interactions -- highlighting how a wide organization of species interactions is fundamentally similar between mutualistic and antagonistic ones \cite{Vazquez2012}. A major direction for future research would be to investigate into more detail these complex interaction networks and to account also for higher-order (not only pairwise) contributions. 

\subsection*{Acknowlegments}
We would like to warmly thank G. Biroli, J.-P. Bouchaud, and S. De Monte for stimulating discussions. 
We also thank M. Baity-Jesi and S. Grigolon for a careful reading of the manuscript.\\
G.G.L. acknowledges the hospitality of the lab. Matière et Systèmes Complexes during her internship. 

\vspace{1cm}

\appendix
\begin{widetext}
\section{The solution in the replica formalism}
\label{app:solution}

We consider the Hamiltonian:
\begin{equation}
	H=\sum_i V_i(N_i)+\sum_{i<j}\alpha_{ij}N_iN_j+(T-m)\sum_i \ln(N_i)
\end{equation}

The disorder average of the free energy can be computed through the replica method \cite{Mezard1986}, a standard tool for fully connected disordered systems based on the identity
\begin{equation}
	-\beta F=\overline{\ln Z}=\lim_{n\to 0}\frac{\ln \overline{Z^n}}{n}
\end{equation}
The quantity on the right hand side can be computed for integer $n$ by introducing $n$ independent replicas of the system, the limit $n\to 0$ is performed through analytic continuation. 
Despite being thermodynamically independent, the replicas are correlated because they are taken at the same realization of the disorder; integrating over the disorder introduces an effective coupling between them. 
%The similarity between their states is measured by the overlap

%Studying the overlap between them allows us to understand the structure of the phase space

Let us thus proceed with the computation of $\overline{Z^n}$:
\begin{equation}
	\overline{Z^n}=\int\prod_{i<j}d\alpha_{ij}\exp\left(-\sum_{i<j}\frac{(\alpha_{ij}-\mu/S)^2}{2\sigma^2/S}\right)\int\prod_{a=1}^n\prod_idN_i^a\exp\left\{-\sum_aH(\{N_i^a\})\right\}
\end{equation}
We can perform the gaussian integration over the $\alpha_{ij}$:
\begin{equation}
% 	\begin{split}
		\overline{Z^n}=\int\prod_{a=1}^n\prod_idN_i^a\exp\Bigg\{\sum_{i<j}\frac{\sigma^2}{2S}\left(\beta \sum_aN_i^aN_j^a\right)^2
		-\beta\sum_a\Bigg[\sum_i(V_i(N_i)+(T-m)\ln(N_i)) + \frac{\mu}{S}\sum_{i<j}N_i^aN_j^a \Bigg] \Bigg\}
% 	\end{split}
\end{equation}
We perform a Hubbard-Stratonovich transformation, introducing the variables $Q_{ab}$, $H_a$ to decouple the abundances of different species; this results in a coupling between different replicas. The replicated partition function can be written as:
\begin{equation}
	\overline{Z^n}=\int\prod_{a\leq b}dQ_{ab}\prod_adH_a\exp\left\{S\mathcal{A}\left(\{Q_{ab}, H_a\}\right)\right\}
\end{equation}
in terms of an action $\mathcal{A}$
\begin{equation}
	\label{eq:action}
		\mathcal{A}\left(\{Q_{ab}, H_a\}\right)
		=-\frac{1}{2}\sigma^2\beta^2\left(\sum_{a<b}Q_{ab}^2+\frac{1}{2}\sum_aQ_{aa}^2\right)+\frac{\beta\mu}{2}\sum_aH_a^2+\frac{1}{S}\sum_i\ln Z_i(\{Q_{ab}, H_a\})
\end{equation}
that contains an effective partition function for the abundances of one species in the different replicas, at fixed values of $Q_{ab}$ and $H_a$:
\begin{equation}
	Z_i=\int\prod_adN_i^a\exp\left\{-\beta H_{eff}(\{N_i^a\},\{Q_{ab}, H_a\})\right\}
\end{equation}
\begin{equation}
	\begin{split}
		H_{eff}=-\beta\sigma^2\left(\sum_{a<b}N_i^aN_i^bQ_{ab}+\frac{1}{2}\sum_a(N_i^a)^2Q_{aa}\right)+\\+\sum_a\bigg(\mu H_aN_i^a+V(N_i^a)+(T-m)\ln N_i^a\bigg)
	\end{split}
\end{equation}

In the $S\to\infty$ limit only the values of $Q_{ab}$ and $H_a$ that extremize the action will contribute\footnote{$H_a$ would actually need to be integrated over the imaginary axis; deforming the integration contour and using the method of steepest descent would then yield the stated result.}:
\begin{equation}
	\frac{\ln \overline{Z^n}}{S}=\mathcal{A}(\{Q_{ab}^*, H_a^*\})
\end{equation}
From the stationarity conditions we obtain:
\begin{align}
	\label{eq:selfconsab}
\begin{aligned}
	\frac{\partial\mathcal{A}}{\partial Q_{ab}}=0\\
	\frac{\partial\mathcal{A}}{\partial H_{a}}=0
\end{aligned}
	\qquad\longrightarrow\qquad
	\begin{aligned}
		Q^*_{ab}=
		\frac{1}{S}\sum_i\langle N_i^{a} N_i^{b}\rangle\\
		H^*_{a}=
		\frac{1}{S}\sum_i\langle N_i^{a} \rangle
	\end{aligned}
\end{align}
The brackets indicate thermal averages under the hamiltonian $H_{eff}$. Since $H_{eff}$ depends on $Q$ and $H$, these equations represent some self-consistency conditions. 

As it is the case for spin glasses systems, we see that $Q_{ab}$ represents the overlap matrix between replicas, and $H_a$ the field. 

\subsection{The replica symmetric solution}
We now need to make an ansatz on the solution. This will be then checked for stability by computing the Hessian of the action, and in particular the first eigenvalue that becomes zero, the replicon \cite{Almeida1978}.
Since the action is symmetric under exchange of replica indices, the first ansatz we make is that the solution respects this symmetry. In this case all the fields must be equal and the overlap can only take two different values: $q_0$ for the overlap between different replicas, and $q_d>q_0$ for the self-overlap.
\begin{align}
	\begin{aligned}
		Q_{ab}=q_0\\
		Q_{aa}=q_d\\
		H_a=h
	\end{aligned}
	&&
	\begin{aligned}
		a\neq b\\
		a=b\\
		\forall a
	\end{aligned}
\end{align}

If there were multiple equilibria, two distinct replicas would have a different overlap according to whether they were in the same state or not.
Since here distinct replicas always have overlap $q_0$, this ansatz corresponds to assuming that there is a unique equilibrium state.
The "size" of this state is characterized by the overlap between different replicas $q_0$: if it is large ($q_0\lesssim q_d$) the configurations in the same state are very similar, so that the state is very localized in phase space, if it is small ($q_0\ll q_d$) the state is very wide. 
We can insert it in the action and in the effective Hamiltonian:
\begin{equation}
	\mathcal{A}(q_d,q_0,h)=-\frac{1}{2}\sigma^2\beta^2\left(\frac{n(n-1)}{2}q_0^2+\frac{n}{2}q_d^2\right)+\frac{\beta\mu n}{2}h^2+\frac{1}{S}\sum_i\ln Z_{i}(q_d, q_0, h)
\end{equation}
\begin{equation}
		H_{eff}=-\frac{\beta\sigma^2}{2}\left(q_0\left(\sum_{a}N_i^a\right)^2+(q_d-q_0)\sum_a(N_i^a)^2\right)+\sum_a\left(\mu hN_i^a+V(N_i^a)+(T-m)\ln N_i^a\right)
\end{equation}
To decouple the different replicas we introduce a Gaussian integration over an auxiliary variable $z_i$, obtaining:
\begin{equation}
	Z_i=\int_{-\infty}^{+\infty}\frac{dz_i}{\sqrt{2\pi}}e^{-z_i^2}\left[\int dN_i\exp\left\{-\beta H_{RS}(N_i; q_d, q_0, h;z)\right\}\right]^n
\end{equation}
\begin{equation}
	H_{RS}=-\beta\sigma^2\frac{q_d-q_0}{2}N_i^2+(\mu h-z\sqrt{q_0}
	\sigma)N_i+V(N_i)+(T-m)\ln N_i
\end{equation}

One of these terms is proportional to a fluctuating field $z$, that accounts for the randomness of the interactions.

Considering again the stationarity condition, performing the analytic continuation for $n \to 0$ and taking into account the fact that after the integration over the fluctuating field $z$ all species are equivalent, the self consistent equations (\ref{eq:selfconsab}) become:
\begin{align}
	\begin{aligned}
	\label{eq:selfconsa}h &=
		\int \mathcal{D}z \left(\frac{\int_0^\infty dN e^{-\beta H_{RS}(q_0, q_d, h, z)}N}{\int_0^\infty dN e^{-\beta H_{RS}(q_0, q_d, h, z)}}\right)= \overline{\langle N \rangle} \\
	q_d & =\int \mathcal{D}z \left(\frac{\int_0^\infty dN e^{-\beta H_{RS}(q_0, q_d, h, z)}N^2}{\int_0^\infty dN e^{-\beta H_{RS}(q_0, q_d, h, z)}}\right) = \overline{\langle N^2 \rangle}\\
	q_0 &=\int \mathcal{D}z \left(\frac{\int_0^\infty dN e^{-\beta H_{RS}(q_0, q_d, h, z)}N}{\int_0^\infty dN e^{-\beta H_{RS}(q_0, q_d, h, z)}}\right)^2= \overline{\langle N \rangle^2} 
	\end{aligned}
\end{align}
where we used the calligraphic notation for the Gaussian integral in $z$: $\int\mathcal{D}z\equiv\int_{-\infty}^{\infty}\frac{dz}{\sqrt{2\pi}}e^{-z^2/2}$.
The brackets indicate the average over the Boltzmann distribution for $N$ under the effective Hamiltonian $H_{RS}$, while the overbar the average over the disorder, represented by $z$.
These averages coincide with thermal and disorder average for a single species abundance.

If we manage to solve the equations (\ref{eq:selfconsa}) we can go back and obtain the free energy as:
\begin{equation}
	f=\frac{F}{S}=-\lim_{n\to0}\frac{\ln \overline{Z^n}}{n \beta S}= - \lim_{n\to 0}\frac{\mathcal{A}(h, q_0, q_d)}{\beta n}
\end{equation}

Nevertheless just knowing if the system is in the replica symmetric phase allows us to understand several physical properties. 
For example, we have seen that the replica symmetric phase corresponds to a unique stable equilibrium, therefore we already know that the equilibrium state will not depend on the assembly history, and we expect the system to relax to it and the fluctuation-dissipation theorem to hold \cite{Cugliandolo2004, Biroli2005, DeDominicis2006}.

\subsubsection{The replicon eigenvalue}

The replica symmetric solution we have considered so far is only correct if it corresponds to a maximum of the action in eq. (\ref{eq:action}). 
To see whether this is the case we can study the harmonic fluctuations around the replica symmetric solution by differentiating the action to second order with respect to the overlap matrix:
\begin{equation}
	\mathcal{M}_{ab,cd}=-\frac{\partial ^2\mathcal{A}}{\partial Q_{ab}\partial Q_{cd}}=\beta^2\sigma^2\left(\delta_{(ab),(cd)}-\beta^2\sigma^2\overline{\langle N^aN^b,N^cN^d\rangle_c}\right)
\end{equation}
$\langle\cdot\rangle_c$ indicates the connected part of the correlator.
If all the eigenvalues of this stability matrix are strictly positive, the fluctuations are finite and the replica symmetric solution is stable.

Thanks to replica symmetry, there are only three independent matrix entries, depending on how many replica indices appear twice: 
\begin{align}
	\begin{aligned}
		\mathcal{M}_{ab,cd}&=-\beta^4\sigma^4\left(\overline{\langle N^aN^bN^cN^d\rangle}-\overline{\langle N^aN^b\rangle^2}\right)\\
		\mathcal{M}_{ab,ac}&=-\beta^4\sigma^4\left(\overline{\langle (N^a)^2N^bN^c\rangle}-\overline{\langle N^aN^b\rangle^2}\right)\\
		\mathcal{M}_{ab,ab}&=\beta^2\sigma^2-\beta^4\sigma^4\left(\overline{\langle (N^a)^2(N^b)^2\rangle}-\overline{\langle N^aN^b\rangle^2}\right)
	\end{aligned}
\end{align}
The stability matrix can be explicitly diagonalized \cite{Almeida1978}. The first eigenvalue to become negative is the replicon, that has the general form:
\begin{equation}
	\lambda = \mathcal{M}_{ab,cd} - 2 \mathcal{M}_{ab,ac} + \mathcal{M}_{ab,ab}
\end{equation}
In the replica symmetric phase, taking the $n\to 0$ limit, we obtain:
\begin{equation}
	\lambda = (\beta \sigma)^2\left(1-(\beta \sigma)^2\overline{(\langle N^2 \rangle-\langle N \rangle^2)^2}\right)
\end{equation}

\subsubsection{Zero temperature, finite size}
\label{app:finitesize}

We can rewrite the effective action at first order in $n$ as:
\begin{equation}
	\mathcal{A}(q_d,q_0,h)=n\beta\left(-\frac{1}{2}\sigma^2q_0\Delta q+\frac{1}{2}\mu h^2+\frac{1}{S\beta n}\sum_i\ln Z_{i}(q_d, q_0, h)\right)
\end{equation}
The quantity in parentheses has a finite limit for $n\to 0$, $\beta\to \infty$.
Because of the $\beta$ factor, we can recover the saddle point approximation by taking the limit $\beta\to \infty$ instead of $S\to \infty$.

\subsection{The replica symmetry broken phase}

When the replicon eigenvalue computed around the replica symmetric solution is negative, replica symmetry is broken and multiple equilibria appear. Following Parisi's scheme \cite{Mezard1986}, we consider a more structured ansatz: replicas are divided into $n/m$ groups of $m$ replicas; the overlap matrix assumes three different (decreasing) values according to whether we are considering the self-overlap, the overlap between replicas in the same group or the overlap between replicas in different groups.

We can then compute again the replicon around the new solution (one-step replica symmetry broken), and if it is still negative we can iterate the procedure, dividing each group in sub-groups. 
At each step the solution is characterized by one more overlap parameter. If an infinite number of steps are required we say the solution displays a full replica symmetry breaking. In this case the replicon approaches zero in the limit of infinite number of steps, so that the solution is marginally stable.

\section{Numerical and analytical results at zero temperature}
\label{numerics}

\begin{figure}[htbp]
	\centering
	\begin{subfigure}[b]{0.44\textwidth}   
		\centering 
		\includegraphics[width=\textwidth]{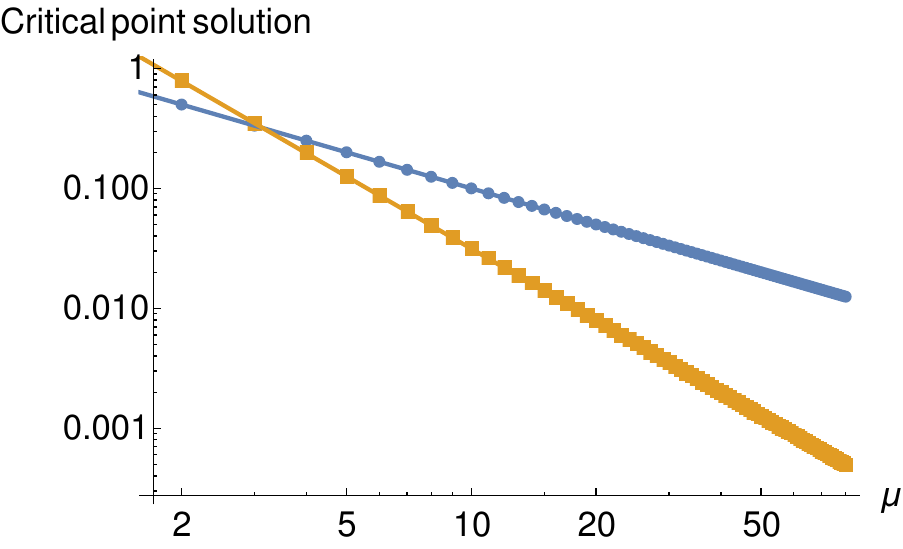}
		\caption[]%
		{\small } 
		\label{fig:hq0sigmac}
	\end{subfigure}
	\hfill
	\begin{subfigure}[b]{0.44\textwidth}   
			\centering
		\includegraphics[width=\textwidth]{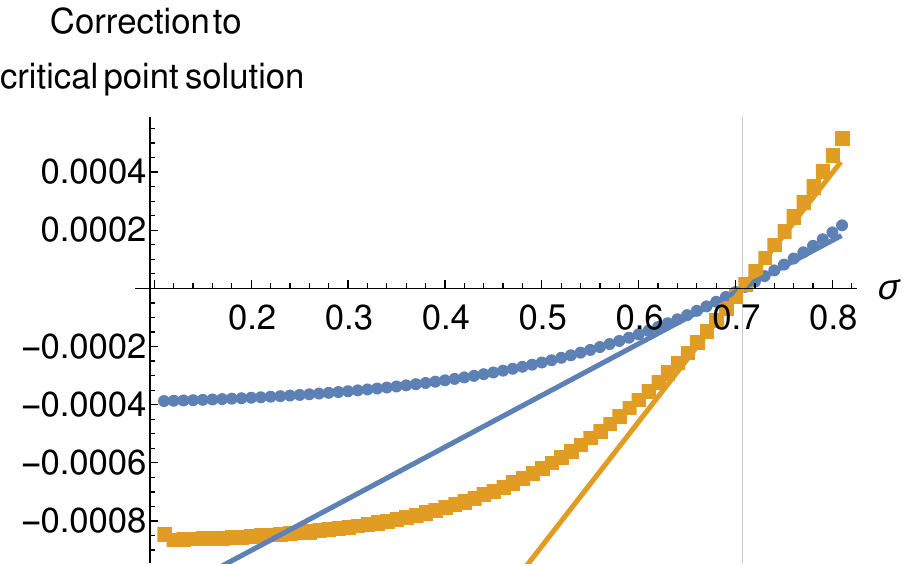}
		\caption[]%
		{{\small }}    
		\label{fig:hq0corr}
	\end{subfigure}
	\caption{{\small Comparison between numerical (dots) and analytical (continuous lines) results at zero temperature. Left: $\overline{\langle N \rangle}=h$ (blue) and $q_0$ (orange) for $\sigma=\sigma_c$. Right: corrections to $h$ (blue) and $q_0$ (orange) to the first order in $\sigma-\sigma_c$}}
	\label{fig:hq0app}
\end{figure}

%\begin{figure}[htbp]
%	\centering
%	\begin{subfigure}[b]{0.44\textwidth}   
%		\centering 
%		\includegraphics[width=\textwidth]{hsigmac.pdf}		\caption[]%\begin{subfigure}[b]{0.44\textwidth}   
	%	\centering
	%	\includegraphics[width=\textwidth]{hcorrection.pdf}
	%	\caption[]%
	%	{{\small }}    
	%	\label{fig:hq0corr}
	%\end{subfigure}
%	\caption{{\small Comparison between numerical (dots) and analytical (continuous lines) results at zero temperature. Left: $\overline{\langle N \rangle}=h$ (blue) and $q_0$ (orange) for $\sigma=\sigma_c$. Right: corrections to $h$ (blue) and $q_0$ (orange) to the first order in $\sigma-\sigma_c$}}
%	\label{fig:hq0app}
%\end{figure}

\section{Large $\mu$ expansion at finite temperature}
\label{app:finitetemperatureexp}
We have obtained the quadratic effective Hamiltonian: 
\begin{equation}
	H_{eff}=\frac{d}{2}(N-N_0)^2
\end{equation}
where $N_0$ is linear in $z$ and $d$ is a constant:
\begin{align}
	d=1-\sigma^2\beta(q_d-q_0)&&
	N_0=-\sqrt{\beta}\frac{\mu h -k-z\sigma\sqrt{q_0}}{\sqrt{2d}}
\end{align}
We can explicitly compute the integrals over $N$ and express them in terms of $N_0$ and $d$. Inserting the results into the self consistent equations (\ref{eq:selfcons1}) we obtain:
\begin{align}
	\begin{aligned}
	h=&\sqrt{\frac{2}{\beta d}}\int \mathcal{D}z \left(\frac{e^{-N_0^2}}{\sqrt{\pi} \erfc(-N_0)}+N_0\right)\\
	q_d=&\frac{2}{\beta d}\int \mathcal{D}z \left(N_0\frac{e^{-N_0^2}}{\sqrt{\pi} \erfc(-N_0)}+(1/2+N_0^2)\right)\\
	q_0=&\frac{2}{\beta d}\int \mathcal{D}z \left(\frac{e^{-N_0^2}}{\sqrt{\pi} \erfc(-N_0)}+N_0\right)^2
	\end{aligned}
\end{align}

As suggested by the numerical results, we assume that $h$, $q_0$ and $q_d$ tend to 0 in the limit of large $\mu$ and that $h$ goes to 0 more slowly than $1/\mu$. These assumptions will be checked for consistency at the end of the computation.
If they hold the constant term in $N_0$ tends to minus infinity, while the factor that multiplies $z$ is small.
Because of the Gaussian weight in $\mathcal{D}z$ (and as it can be checked numerically) only values of $z$ of order 1 contribute to the integral, so that the argument of the complementary error function is large and we can use the asymptotic expansion:
\begin{align}
	\erfc(x)=\frac{e^{-x^2}}{\sqrt{\pi}x}\sum_{n=0}^\infty(-1)^n\frac{(2n-1)!!}{(2x^2)^n}\\
	\frac{e^{-N_0^2}}{\sqrt{\pi}\erfc(-N_0)}=-N_0-\frac{1}{2N_0}+\frac{1}{2N_0^3}+O\left(1/N_0^5\right)
\end{align}
Inserting the first non vanishing term of this expansion into the equation for $h$ we obtain an integrand with only fundamental functions of $z$: 
\begin{equation}
	h\approx-\sqrt{\frac{2}{\beta d}}\int \mathcal{D}z\frac{1}{2N_0}=\frac{1}{\beta }\int dz \frac{e^{-z^2}}{\sqrt{2\pi}}\frac{1}{\mu h - k -z \sigma\sqrt{q_0}}
\end{equation}
The integral has now a divergence in $z=-(\mu h-k)/(\sigma \sqrt{q_0})$ that was not present in the original expression and that is due to the failing of the asymptotic expansion for $N_0 \sim 0$. Remembering that we expect the main contribution to the integral to be given by $z\sim O(1)$ and that $\mu h -k\gg\sigma\sqrt{q_0}$, we expand
\begin{equation}
h\approx\frac{1}{\beta }\int dz \frac{e^{-z^2}}{\sqrt{2\pi}}\frac{1}{\mu h - k -z \sigma\sqrt{q_0}}
	\approx\frac{1}{\beta}\int dz \frac{e^{-z^2}}{\sqrt{2\pi}}\left(\frac{1}{\mu h -k}-\frac{\sigma \sqrt{q_0}z}{(\mu h -k)^2}+...\right)
\end{equation}
The integrals in $z$ are now immediate; considering only the dominant order we obtain a simple equation for $h$:
\begin{equation}
	h\approx\frac{1}{\beta(\mu h-k)}
\end{equation}

Similarly,
	\begin{align}
		\begin{aligned}
		q_0\approx\frac{2}{\beta d}\int \mathcal{D}z \frac{1}{4c^2}\approx\frac{1}{\beta^2(\mu h-k)^2}=h^2\\
	q_d\approx\frac{2}{\beta d}\int \mathcal{D}z \frac{1}{2c^2}\approx\frac{2}{\beta^2(\mu h-k)^2}=2h^2
	\end{aligned}
\end{align}
Solving the second order equation for $h$, expanding the solution at the dominant order in $1/\mu$ and plugging it into the equations for $q_0$ and $q_d$, we get the closed form results:
\begin{align}
	\begin{aligned}
		\label{eq:hq0qdanapp}	h&\approx\frac{k+\sqrt{k^2+4\mu /\beta}}{2\mu}\approx\frac{1}{\sqrt{\beta\mu}}\\
		q_0&\approx\frac{1}{\beta \mu}\\
		q_d&\approx\frac{2}{\beta \mu}
	\end{aligned}
\end{align}
In the last expansion for $h$ we have neglected higher powers of $\sqrt{\frac{k^2\beta}{\mu}}$, therefore we expect the results to be correct only when $\mu\gg\beta$.
\subsection{Replicon eigenvalue}
\label{app:replicontfinite}

% \begin{figure}[htbp]
% \centering
% \includegraphics[width=0.65\textwidth]{repliconfiniteT}
% 	\caption{\small{Numerical (dots) and analytic (continuous line) results for the replicon eigenvalue. $\beta=1$, $\sigma=1/5$.}}
% \label{fig:csi}
% \end{figure}
We can compute the replicon eigenvalue:
\begin{equation}
	\lambda = (\beta \sigma)^2\left(1-(\beta \sigma)^2\overline{(\langle N^2 \rangle-\langle N \rangle^2)^2}\right)
\end{equation}
The quantity under overbar can be computed in a similar way to what was done in the previous section, giving:
\begin{equation}
	\begin{split}
		\overline{(\langle N^2 \rangle-\langle N \rangle^2)^2}&=\frac{4}{(\beta d)^2}\int \mathcal{D}z \left\{\frac{1}{2}-N_0\frac{e^{-N_0^2}}{\sqrt{\pi} \erfc(-N_0)}-\left(\frac{e^{-N_0^2}}{\sqrt{\pi} \erfc(-N_0)}\right)^2\right\}^2\approx\\
		&\approx\frac{1}{4(\beta db)^2}=h^4=\frac{1}{\beta^2\mu^2}
	\end{split}
\end{equation}
We thus obtain:
\begin{equation}
	\label{eq:repliconmugrandeapp}
	\lambda = (\beta \sigma)^2\left(1-\frac{\sigma^2}{\mu^2}\right)>0
\end{equation}
\end{widetext}

\bibliographystyle{unsrt}
\bibliography{ZoteroLibrary}

\begin{thebibliography}{10}

\bibitem{Vazquez2012}
Diego~P V{\'a}zquez, Silvia~B Lom{\'a}scolo, M~Bel{\'e}n Maldonado, Natacha~P
  Chacoff, Jimena Dorado, Erica~L Stevani, and Nydia~L Vitale.
\newblock The strength of plant--pollinator interactions.
\newblock {\em Ecology}, 93(4):719--725, 2012.

\bibitem{Bunin2021}
Guy Bunin.
\newblock Directionality and community-level selection.
\newblock {\em Oikos}, 130(4):489--500, 2021.

\bibitem{Tilman2004niche}
David Tilman.
\newblock Niche tradeoffs, neutrality, and community structure: a stochastic
  theory of resource competition, invasion, and community assembly.
\newblock {\em Proceedings of the National Academy of Sciences},
  101(30):10854--10861, 2004.

\bibitem{gupta2021}
Deepak Gupta, Stefano Garlaschi, Samir Suweis, Sandro Azaele, and Amos Maritan.
\newblock An effective resource-competition model for species coexistence.
\newblock {\em arXiv preprint arXiv:2104.01256}, 2021.

\bibitem{Hardin1960}
G.~Hardin.
\newblock The {{Competitive Exclusion Principle}}.
\newblock {\em Science}, 131(3409):1292--1297, April 1960.

\bibitem{azaele2016}
Sandro Azaele, Samir Suweis, Jacopo Grilli, Igor Volkov, Jayanth~R. Banavar,
  and Amos Maritan.
\newblock Statistical mechanics of ecological systems: {{Neutral}} theory and
  beyond.
\newblock {\em Reviews of Modern Physics}, 88(3):035003, July 2016.

\bibitem{Arnoldi2018}
J-F Arnoldi, Azenor Bideault, Michel Loreau, and Bart Haegeman.
\newblock How ecosystems recover from pulse perturbations: A theory of short-to
  long-term responses.
\newblock {\em Journal of theoretical biology}, 436:79--92, 2018.

\bibitem{Roy2020}
Felix Roy, Matthieu Barbier, Giulio Biroli, and Guy Bunin.
\newblock Complex interactions can create persistent fluctuations in
  high-diversity ecosystems.
\newblock {\em PLoS Computational Biology}, 16(5):1--14, 2020.

\bibitem{Altieri2021}
Ada Altieri and Giulio Biroli.
\newblock Effects of intraspecific cooperative interactions in large
  ecosystems.
\newblock {\em arXiv preprint arXiv:2105.04519}, 2021.

\bibitem{Zaoli2021}
Silvia Zaoli and Jacopo Grilli.
\newblock A macroecological description of alternative stable states reproduces
  intra-and inter-host variability of gut microbiome.
\newblock {\em bioRxiv}, 2021.

\bibitem{Beninca2008}
Elisa Beninc{\'a}, Jef Huisman, Reinhard Heerkloss, Klaus~D. J{\"o}hnk, Pedro
  Branco, Egbert~H. Van~Nes, Marten Scheffer, and Stephen~P. Ellner.
\newblock Chaos in a long-term experiment with a plankton community.
\newblock {\em Nature}, 451(7180):822--825, 2008.

\bibitem{Pearce2020}
Michael~T Pearce, Atish Agarwala, and Daniel~S Fisher.
\newblock Stabilization of extensive fine-scale diversity by ecologically
  driven spatiotemporal chaos.
\newblock {\em Proceedings of the National Academy of Sciences},
  117(25):14572--14583, 2020.

\bibitem{Tilman1982}
David Tilman.
\newblock {\em Resource competition and community structure}.
\newblock Princeton university press, 1982.

\bibitem{van2015}
Joost~HJ van Opheusden, Lia Hemerik, Mieke van Opheusden, and Wopke van~der
  Werf.
\newblock Competition for resources: complicated dynamics in the simple tilman
  model.
\newblock {\em SpringerPlus}, 4(1):474, 2015.

\bibitem{Gibbs2018}
Theo Gibbs, Jacopo Grilli, Tim Rogers, and Stefano Allesina.
\newblock Effect of population abundances on the stability of large random
  ecosystems.
\newblock {\em Physical Review E}, 98(2):022410, 2018.

\bibitem{Fisher2014}
Charles~K. Fisher and Pankaj Mehta.
\newblock The transition between the niche and neutral regimes in ecology.
\newblock {\em Proceedings of the National Academy of Sciences of the United
  States of America}, 111(36):13111--13116, 2014.

\bibitem{Barbier2018}
Matthieu Barbier, Jean~Fran{\c c}ois Arnoldi, Guy Bunin, and Michel Loreau.
\newblock Generic assembly patterns in complex ecological communities.
\newblock {\em Proceedings of the National Academy of Sciences of the United
  States of America}, 115(9):2156--2161, 2018.

\bibitem{Sidhom2020}
Laura Sidhom and Tobias Galla.
\newblock Ecological communities from random generalized lotka-volterra
  dynamics with nonlinear feedback.
\newblock {\em Physical Review E}, 101(3):032101, 2020.

\bibitem{Bunin2016}
Guy Bunin.
\newblock Interaction patterns and diversity in assembled ecological
  communities.
\newblock July 2016.

\bibitem{bunin2017}
Guy Bunin.
\newblock Ecological communities with {{Lotka}}-{{Volterra}} dynamics.
\newblock {\em Physical Review E}, 95(4):1--8, 2017.

\bibitem{Lotka1920}
Alfred~J. Lotka.
\newblock Analytical {{Note}} on {{Certain Rhythmic Relations}} in {{Organic
  Systems}}.
\newblock {\em Proceedings of the National Academy of Sciences}, 6(7):410--415,
  July 1920.

\bibitem{volterra1926}
V~Volterra.
\newblock Variazioni e fluttuazioni del numero d'individui in specie animali
  conviventi.
\newblock {\em Memoria della regia accademia nazionale del lincei ser.},
  62:31--113, 1926.

\bibitem{MacArthur1955}
Robert MacArthur.
\newblock Fluctuations of {{Animal Populations}} and a {{Measure}} of
  {{Community Stability}}.
\newblock {\em Ecological Society of America}, 36(3):533--536, 1955.

\bibitem{May2007}
Robert May and Angela~R. McLean.
\newblock {\em Theoretical {{Ecology}}}.
\newblock {Oxford University Press}, February 2007.

\bibitem{moran2019}
Jos{\'e} Moran and Jean-Philippe Bouchaud.
\newblock May's instability in large economies.
\newblock {\em Physical Review E}, 100(3):032307, September 2019.

\bibitem{Goodwin1990}
Richard~M. Goodwin.
\newblock {\em Chaotic {{Economic Dynamics}}}.
\newblock {Oxford University Press}, November 1990.

\bibitem{Behn1992}
Ulrich Behn, J.~Leo {van Hemmen}, and Bernhard Sulzer.
\newblock Memory {{B Cells Stabilize Cycles}} in a {{Repressive Network}}.
\newblock In {\em Theoretical and {{Experimental Insights}} into
  {{Immunology}}}, pages 249--260. {Springer Berlin Heidelberg}, {Berlin,
  Heidelberg}, 1992.

\bibitem{Bomze1995}
Immanuel~M. Bomze.
\newblock Lotka-{{Volterra}} equation and replicator dynamics: New issues in
  classification.
\newblock {\em Biological Cybernetics}, 72(5):447--453, 1995.

\bibitem{hu2021}
Jiliang Hu, Daniel~R. Amor, Matthieu Barbier, Guy Bunin, and Jeff Gore.
\newblock Emergent phases of ecological diversity and dynamics mapped in
  microcosms.
\newblock Preprint, {Biophysics}, October 2021.

\bibitem{biroli2018}
Giulio Biroli, Guy Bunin, and Chiara Cammarota.
\newblock Marginally stable equilibria in critical ecosystems.
\newblock {\em New Journal of Physics}, 20(8):083051, August 2018.

\bibitem{altieri2021a}
Ada Altieri, Felix Roy, Chiara Cammarota, and Giulio Biroli.
\newblock Properties of {{Equilibria}} and {{Glassy Phases}} of the {{Random
  Lotka}}-{{Volterra Model}} with {{Demographic Noise}}.
\newblock {\em Physical Review Letters}, 126(25):258301, June 2021.

\bibitem{Kessler2015}
David~A Kessler and Nadav~M Shnerb.
\newblock Generalized model of island biodiversity.
\newblock {\em Physical Review E}, 91(4):042705, April 2015.

\bibitem{Baron2019}
Matthieu Baron, Silvia~De Monte, and Giulio Biroli.
\newblock Role of the interaction network in large interacting ecosystems,
  {{Internship}} report, 2019.

\bibitem{De2015}
Colomban De~Vargas, St{\'e}phane Audic, Nicolas Henry, Johan Decelle,
  Fr{\'e}d{\'e}ric Mah{\'e}, Ramiro Logares, Enrique Lara, C{\'e}dric Berney,
  Noan Le~Bescot, Ian Probert, et~al.
\newblock Eukaryotic plankton diversity in the sunlit ocean.
\newblock {\em Science}, 348(6237), 2015.

\bibitem{SerGiacomi2018}
Enrico {Ser-Giacomi}, Lucie Zinger, Shruti Malviya, Colomban De~Vargas, Eric
  Karsenti, Chris Bowler, and Silvia De~Monte.
\newblock Ubiquitous abundance distribution of non-dominant plankton across the
  global ocean.
\newblock {\em Nature Ecology \& Evolution}, 2(8):1243--1249, August 2018.

\bibitem{ratzke2020}
Christoph Ratzke, Julien Barrere, and Jeff Gore.
\newblock Strength of species interactions determines biodiversity and
  stability in microbial communities.
\newblock {\em Nature Ecology \& Evolution}, 4(3):376--383, March 2020.

\bibitem{bauer2018}
Maria~A. Bauer, Katharina Kainz, Didac {Carmona-Gutierrez}, and Frank Madeo.
\newblock Microbial wars: Competition in ecological niches and within the
  microbiome.
\newblock {\em Microbial Cell}, 5(5):215--219, May 2018.

\bibitem{May1972}
Robert~M. May.
\newblock Will a {{Large Complex System}} be {{Stable}}?
\newblock {\em Nature}, 238(5364):413--414, August 1972.

\bibitem{MacArthur2016}
Robert~H MacArthur and Edward~O Wilson.
\newblock {\em The theory of island biogeography}.
\newblock Princeton university press, 2016.

\bibitem{Rieger1989}
H.~Rieger.
\newblock Solvable model of a complex ecosystem with randomly interacting
  species.
\newblock {\em Journal of Physics A: Mathematical and General},
  22(17):3447--3460, 1989.

\bibitem{Galla2018}
Tobias Galla.
\newblock Dynamically evolved community size and stability of random
  {{Lotka}}-{{Volterra}} ecosystems(a).
\newblock {\em Epl}, 123(4):1--13, 2018.

\bibitem{Blanchard2015}
Andrew~E Blanchard and Ting Lu.
\newblock Bacterial social interactions drive the emergence of differential
  spatial colony structures.
\newblock {\em BMC systems biology}, 9(1):1--13, 2015.

\bibitem{Carr2019}
Alex Carr, Christian Diener, Nitin~S Baliga, and Sean~M Gibbons.
\newblock Use and abuse of correlation analyses in microbial ecology.
\newblock {\em The ISME journal}, 13(11):2647--2655, 2019.

\bibitem{fried2016}
Yael Fried, David~A. Kessler, and Nadav~M. Shnerb.
\newblock Communities as cliques.
\newblock {\em Scientific Reports}, 6(1):35648, October 2016.

\bibitem{Mezard1986}
M~Mezard, G~Parisi, and M~Virasoro.
\newblock {\em Spin {{Glass Theory}} and {{Beyond}}: {{An Introduction}} to the
  {{Replica Method}} and {{Its Applications}}}, volume~9 of {\em World
  {{Scientific Lecture Notes}} in {{Physics}}}.
\newblock {WORLD SCIENTIFIC}, November 1986.

\bibitem{Almeida1978}
J~R~L de~Almeida and D~J Thouless.
\newblock Stability of the {{Sherrington}}-{{Kirkpatrick}} solution of a spin
  glass model.
\newblock {\em Journal of Physics A: Mathematical and General}, 11(5):983--990,
  May 1978.

\bibitem{DeDominicis2006}
Cirano De~Dominicis and Irene Giardina.
\newblock {\em Random {{Fields}} and {{Spin Glasses}}}.
\newblock {Cambridge University Press}, {Cambridge}, September 2006.

\bibitem{Diederich1989}
S.~Diederich and M.~Opper.
\newblock Replicators with random interactions: {{A}} solvable model.
\newblock {\em Physical Review A}, 39(8):4333--4336, 1989.

\bibitem{Biscari1995}
Paolo Biscari and G~Parisi.
\newblock Replica symmetry breaking in the random replicant model.
\newblock {\em Journal of Physics A: Mathematical and General}, 28(17):4697,
  1995.

\bibitem{Tikhonov2017}
Mikhail Tikhonov and Remi Monasson.
\newblock Collective {{Phase}} in {{Resource Competition}} in a {{Highly
  Diverse Ecosystem}}.
\newblock {\em Physical Review Letters}, 118(4):1--5, 2017.

\bibitem{Altieri2019}
Ada Altieri and Silvio Franz.
\newblock Constraint satisfaction mechanisms for marginal stability and
  criticality in large ecosystems.
\newblock {\em Physical Review E}, 99(1):010401, 2019.

\bibitem{Gardner1988}
E.~Gardner and B.~Derrida.
\newblock Optimal storage properties of neural network models.
\newblock {\em Journal of Physics A: General Physics}, 21(1):271--284, 1988.

\bibitem{DeMartino2004}
A~De Martino, M.~Marsili, and I~P{\'e}rez Castillo.
\newblock Statistical mechanics analysis of the equilibria of linear economies.
\newblock {\em Journal of Statistical Mechanics: Theory and Experiment},
  2004(04):P04002, April 2004.

\bibitem{grilli2020}
Jacopo Grilli.
\newblock Macroecological laws describe variation and diversity in microbial
  communities.
\newblock {\em Nature Communications}, 11(1):4743, December 2020.

\bibitem{Dubinkina2019}
Veronika Dubinkina, Yulia Fridman, Parth~Pratim Pandey, and Sergei Maslov.
\newblock Multistability and regime shifts in microbial communities explained
  by competition for essential nutrients.
\newblock {\em Elife}, 8:e49720, 2019.

\bibitem{Cugliandolo2004}
L.~F. Cugliandolo.
\newblock Course 7: {{Dynamics}} of {{Glassy Systems}}.
\newblock In {\em Slow {{Relaxations}} and Nonequilibrium Dynamics in Condensed
  Matter. {{Les Houches}}-{{\'Ecole}} d'{{\'Et\'e}} de {{Physique Theorique}},
  Vol 77.} {Springer, Berlin, Heidelberg}, 2004.

\bibitem{Biroli2005}
Giulio Biroli.
\newblock A crash course on ageing.
\newblock {\em Journal of Statistical Mechanics: Theory and Experiment},
  2005(05):P05014, May 2005.

\end{thebibliography}
\end{document}